\documentclass[fleqn,usenatbib]{mnras}

\usepackage{txfonts}
\usepackage[T1]{fontenc}

\usepackage{color}
\usepackage{float}
\usepackage[pdftex]{graphicx}
\usepackage{epstopdf}
\usepackage{natbib}
 \usepackage{appendix}
\usepackage{wasysym}

\DeclareRobustCommand{\VAN}[3]{#2}
\let\VANthebibliography\thebibliography
\def\thebibliography{\DeclareRobustCommand{\VAN}[3]{##3}\VANthebibliography}

\newcommand{\jwst}[0]{\textit{JWST }}
\newcommand{\hst}[0]{\textit{HST }}
\newcommand{\spitzer}[0]{\textit{Spitzer }}
\newcommand{\herschel}[0]{\textit{Herschel }}

\newcommand{\lsun}{\hbox{L$_\odot$}}

\newcommand{\msun}{\hbox{M$_\odot$}}

\title[Extremely Dusty Galaxies]{Identifying and Characterizing the Most Heavily Dust-Obscured Galaxies at $1 \le z \le 4$}

\author[Martis et al.]{
Nicholas S. Martis,$^{1,2,3}$\thanks{E-mail: nicholas.martis@smu.ca}
Danilo M. Marchesini,$^{3}$ 
Adam Muzzin,$^{4}$
\newauthor{Chris J. Willott,$^{2}$ 
Marcin Sawicki$^{1}$}
\\
$^{1}$Astronomy and Physics Department, St. Mary's University, Halifax, NS \\
${2}$National Research Council of Canada, Herzberg Astronomy \& Astrophysics Research Centre, 5071 West Saanich Road, Victoria, BC, Canada, V9E 2E7 \\
${3}$Physics and Astronomy Department, Tufts University, Medford, MA 02155 \\
${4}$Department of Physics and Astronomy, York University, 4700 Keele St., Toronto, Ontario, Canada, MJ3 1P3
}

\date{Accepted XXX. Received YYY; in original form ZZZ}

\pubyear{2022}

\begin{document}
\label{firstpage}
\pagerange{\pageref{firstpage}--\pageref{lastpage}}
\maketitle

\begin{abstract}
We present 65 extremely dust-obscured galaxies from the UltraVISTA DR3 survey of the COSMOS field at $1<z<4$. In contrast to other studies of dusty galaxies, we select our sample based on dust attenuation measured by UV-MIR spectral energy distribution (SED) modeling that allows for extreme attenuation levels.
We construct our sample by making cuts at $1 \le z \le 4$, A$_V \ge 3$, and log(M$_*/\msun) \ge 10.5$. This method reliably selects galaxies exhibiting independent indicators of significant dust content, including FIR detection rates. We perform panchromatic SED modeling with matched \textit{Herschel} photometry and find stellar and dust properties that differ from typical sub-millimeter galaxy (SMG) samples as well as \mbox{\herschel} sources matched in redshift and stellar mass. Our sources have lower star formation rates and higher A$_V$ than SMGs, but comparable total IR luminosities. Most of our sample falls on or near the star-forming main sequence for this redshift range. Finally, we perform a morphological analysis with GALFIT using the $K_S$-band images and \textit{Hubble} $F814W$ and $F160W$ imaging when available. Typical axis ratios of $\sim 0.4$ suggest disk-like morphology for the majority of our sources, and we note only three apparent merging systems. Our sample generally agrees with the size-mass relation for star-forming galaxies, with a tail extending to smaller sizes. We conclude that the most heavily obscured galaxies in this redshift range share many characteristics with typical star-forming galaxies, forming a population of dusty galaxies that overlaps, but is not encompassed by, those selected through dust emission. 

\end{abstract}


\section{Introduction}
One of the primary goals of extragalactic astronomy is to understand how populations of galaxies observed in the early universe evolve to become the galaxies in the local universe today. As telescope technology has improved, astronomers have consistently uncovered new classes of galaxies that must fit into this puzzle. The first wide and deep surveys in the near-infrared and the launch of the \textit{Spitzer Space Telescope} led to the discovery of a class of objects with very red optical-NIR colors at high redshift. This discovery provided a new window on the high-redshift universe beyond the well-studied Lyman break galaxies. Several simple color selection criteria were developed, each with a designation for their associated populations including Extremely Red objects \citep[EROs;][]{roche03}, IRAC-selected EROs \citep[IEROs, also known as IR EROs;][]{yan04}, and distant red galaxies \citep[DRGs;][]{franx03}. 

Since both passive galaxies and galaxies with substantial dust attenuation satisfy these types of color selections, the next necessary step to understand these objects was to identify reliable ways to distinguish between these two main groups within the overall population of EROs \citep{kong09, fang09, castro-rodriguez12}. Despite the progress, these color selections result in samples with rather broad redshift ranges. With the development of surveys covering the UV-to-NIR with many bands and with extension to 8 $\micron$ in the best cases, it finally became possible to calculate accurate photometric redshifts for large samples of galaxies \citep{kong06, blanc08}. With photometric redshifts, it became possible to classify large samples based on their rest-frame colors, thus providing a clearer view of their stellar population properties. To this end, the rest-frame $U-V$, $V-J$ color-color (hereafter UVJ) diagram has become a common tool when available data are restricted to UV-to-MIR photometry. Though it is much more commonly used to distinguish quiescent and star-forming galaxies, it has been shown capable of selecting specifically dusty star-forming galaxies  as well \citep{spitler14, martis16, fang18}. 

The $\spitzer$ Multiband Imaging Photometer (MIPS) has also seen wide use in selecting dust-obscured galaxies \citep[DOGs in the literature;][]{dey08, bussmann09,riguccini15, oi17}. At $z \gtrsim 1$, the polycyclic aromatic hydrocarbon (PAH) features which trace star formation are shifted into the MIPS 24 $\micron$ band, so in the absence of an active galactic nucleus (AGN), a 24 $\micron$ flux is often directly translated into an obscured star formation rate. This appears to be valid for starbursts, but recent work has shown that dust heating from intermediate age stellar populations can complicate this measurement. \citep{utomo14, leja19,martis19, roebuck19}. In terms of AGN activity, DOGs constitute a heterogeneous population of starbursts and obscured AGN as well as composites of the two. Working in a similar wavelength range as MIPS, the Wide-field Infrared Explorer (\textit{WISE}) satellite conducted an all-sky survey at 22 $\micron$ \citep{wright10}. While shallower than most MIPS observations, the wide coverage did allow for the detection of large numbers of very luminous dusty galaxies, including many at high redshift \citep{bridge13,blain13,tsai13}.

The final major group of dusty galaxies defined by an observational criterion are the sub-millimeter galaxies (SMGs), which were first characterized only slightly earlier \citep{smail97, barger98, blain02}. Significant effort has gone into relating SMGs to the dusty galaxies selected in the NIR and $24 \micron$, a process sometimes hampered by faint or undetected optical counterparts \citep{pope08b, pope08a, bussmann09, kirkpatrick12}.  These galaxies are strongly star-forming, heavily obscured by dust, and frequently host AGN, but their physical origin remains a point of debate \citep{hayward11, hayward12, magnelli12, bethermin15, elbaz18}. Major mergers appear necessary to achieve the star formation rates and IR luminosities of these systems in the local universe, but it is not clear that this is the case at higher redshift.

The relative rarity of such luminous systems means that large survey areas are required to construct statistical samples of them. The launch of the \textit{Herschel Space Observatory} enabled mapping of wide areas of the sky in the 250-500 $\micron$ range with the SPIRE instrument \citep{oliver12,eales10}. When combined with MIR measurements, which can serve the dual purpose of aiding in deblending the FIR images which suffer from beam sizes in the tens of arcseconds, these FIR observations enable the measurement of a galaxy's total IR luminosity. Alongside the SMG nomenclature, there is a literature of (ultra) luminous infrared galaxies, or (U)LIRGS, defined as having L$_{IR} > 10^{11}$L$_\odot$ ($10^{12}$L$_\odot$ for ULIRGS). These sources range from normal star-forming galaxies at high redshift to extreme merger induced starbursts and, like SMGs, often host AGN. Sampling multiple points along the FIR SEDs of galaxies additionally made it possible to constrain the thermal dust temperature in distant galaxies \citep{thomson17,schreiber18}. This measurement may be strengthened by the inclusion of photometry in the 70-160 $\micron$ range from the \textit{Herschel} PACS instrument, which also conducted large extragalactic surveys \citep{lutz11,elbaz11}, albeit not covering as wide an area as the SPIRE surveys. 

From a physical standpoint, each of these classes of IR-selected sources requires its dust to be heated by either excessive star formation or an AGN. Even though they do indeed exhibit significant levels of dust obscuration in the UV-optical to coincide with their dust emission (when the measurement is available), we have shown in \citet{martis19} that a selection in dust emission does not equate to a selection in dust obscuration. Specifically we showed that within a sample of galaxies selected by dust obscuration, the additional requirement of a FIR detection with \textit{Herschel} biases the sample to lower redshift as well as higher stellar masses, star formation rates, and dust luminosities. In principle, a description of the full population of dusty galaxies should account for both groups of galaxies if they are indeed distinct. 

The first step toward this goal is to identify and characterize samples of both types. Many effects might contribute to differences in selections based on dust emission or obscuration, with the most obvious being the sensitivity of mid- and far-IR surveys. Other more subtle effects include the shape of the attenuation curve in the region attenuation is measured, the star-dust geometry, the chemical composition of the dust, and inclination angle of our line of sight toward the galaxy. This work examines the physical properties of a sample of heavily obscured galaxies in order to investigate the causes that lead to their high obscuration levels.

We have identified a set of sources in the UltraVISTA survey with exceptionally red SEDs for which initial modeling implies A$_V > 3$ mag. Primarily, we wish to better understand the nature of these extreme objects, but we also examine the relation between these sources and other selections of dusty galaxies. The bulk of the aforementioned studies focus on selecting dusty galaxies based on either observed colors or IR emission. The present study takes a different approach and identifies a population of galaxies with extreme dust obscuration through the use of SED modeling of UV-to-MIR data. To our knowledge, this is the first study which selects dusty galaxies in this manner. Through comparison with MIR and FIR data as well as UV-to-FIR SED modeling, we show that this technique efficiently selects IR-bright galaxies, providing independent confirmation their dusty nature. This paper will be organized as follows. Section 2 will describe the data used in this study. Section 3 outlines our selection method and SED modeling technique. Section 4 presents our results and our discussion of them. We conclude in Section 5. Throughout this paper, we adopt a \citet{chabrier03} IMF. All magnitudes are in the AB system. We assume a cosmology with $\Omega_\Lambda$ = 0.7, $\Omega_M$ = 0.3, and $H_0$ = 70 km s$^{-1}$Mpc$^{-1}$.

\section{Data}
We select sources from the DR3 of the UltraVISTA $K_S$-band-selected photometric catalog. The DR3 catalog (Muzzin et al. in prep.) was constructed using the same procedure as in \citet{muzzin13a}. The relevant details for the current data release are outlined in \citet{martis19}. In brief, the survey covers 0.7 deg$^2$ in the COSMOS field as deep stripes overlapping the DR1. The catalog includes photometry from the UV to \textit{Spitzer} 8$\micron$ with 49 bands. Sources are selected in the K$_S$-band, which has a point source 5$\sigma$ depth of 25.2 magnitudes. 
In addition, \textit{Spitzer} MIPS 24$\micron$ observations \citep{lefloch09} are included as follows. Briefly, we assume that there are no color gradients in galaxies between the K$_S$-band and the IRAC and MIPS bands. The K$_S$-band is then used as a high-resolution template image to deblend the IRAC and MIPS photometry. Each source extracted from the K$_S$ image is convolved with a kernel derived from bright PSF stars in the K$_S$ and IRAC/MIPS images. The convolved galaxies are then fit as templates in the IRAC and MIPS bands with the total flux left as a free parameter. In this process, all objects in the image are fit simultaneously. Once the template fitting is converged, a “cleaned” image is produced for each object in the catalog by subtracting off all nearby sources \citep[for an example of this process see Figure 1 of][]{wuyts07}. Aperture photometry is then performed on the cleaned image of each source. For the IRAC (MIPS) channels the photometry is performed in a 3" (5") diameter aperture for each object. All UV-NIR fluxes are scaled to total using the ratio of total to aperture fluxes for the K$_S$-band and MIPS fluxes have been converted to total fluxes using an aperture correction factor of 3.7, as listed in the MIPS instrument handbook.

The inclusion of \herschel PACS and SPIRE data used later in this analysis is also identical to that described in \citet{martis19}. We supplement the UltraVISTA data with observations from the \textit{Herschel} PACS Evolutionary Probe  \citep[PEP;][]{lutz11} and \textit{Herschel} Multi-Tiered Extragalactic Survey \citep[HerMES;][]{oliver12, hurley17} surveys. The PEP survey covers most of the UltraVISTA footprint with the PACS 100$\micron$ and 160$\micron$ filters for which the 80\% completeness level is reached at 6.35 and 14.93 mJy (5$\sigma$ calculated from one sigma noise is 7.50 and 16.35 mJy), respectively. HerMES covers the area with the 250\micron, 350\micron, and 500$\micron$ filters at 5$\sigma$ depths of 15.9, 13.3, and 19.1 mJy, respectively. For both surveys we use the source catalogs extracted using MIPS 24$\micron$ priors which also use the \citet{lefloch09} data. For HerMES we use the most recent XID+ catalogs \citep{hurley17}. The positional accuracy of MIPS at 24 $\micron$ is about 2", yielding much more accurate source positions than would be possible with a blind extraction. Sources with at least a 3$\sigma$ detection in any of the Herschel bands are matched to UltraVISTA sources within a matching radius of 1.5" provided the corresponding UltraVISTA source is detected in MIPS. The requirement of a MIPS detection ensures that the \textit{Herschel} sources are being matched to IR-bright Ultravista sources, lowering the probability of false matches that may otherwise occur with the relatively large matching radius. In combination with the careful approach taken to deblend MIPS sources in the UltraVISTA photometry described above, this ensures that we have the most reliable matching of the UltraVISTA and \textit{Herschel} catalogs possible.

\begin{figure*}
    \includegraphics[width=\textwidth]{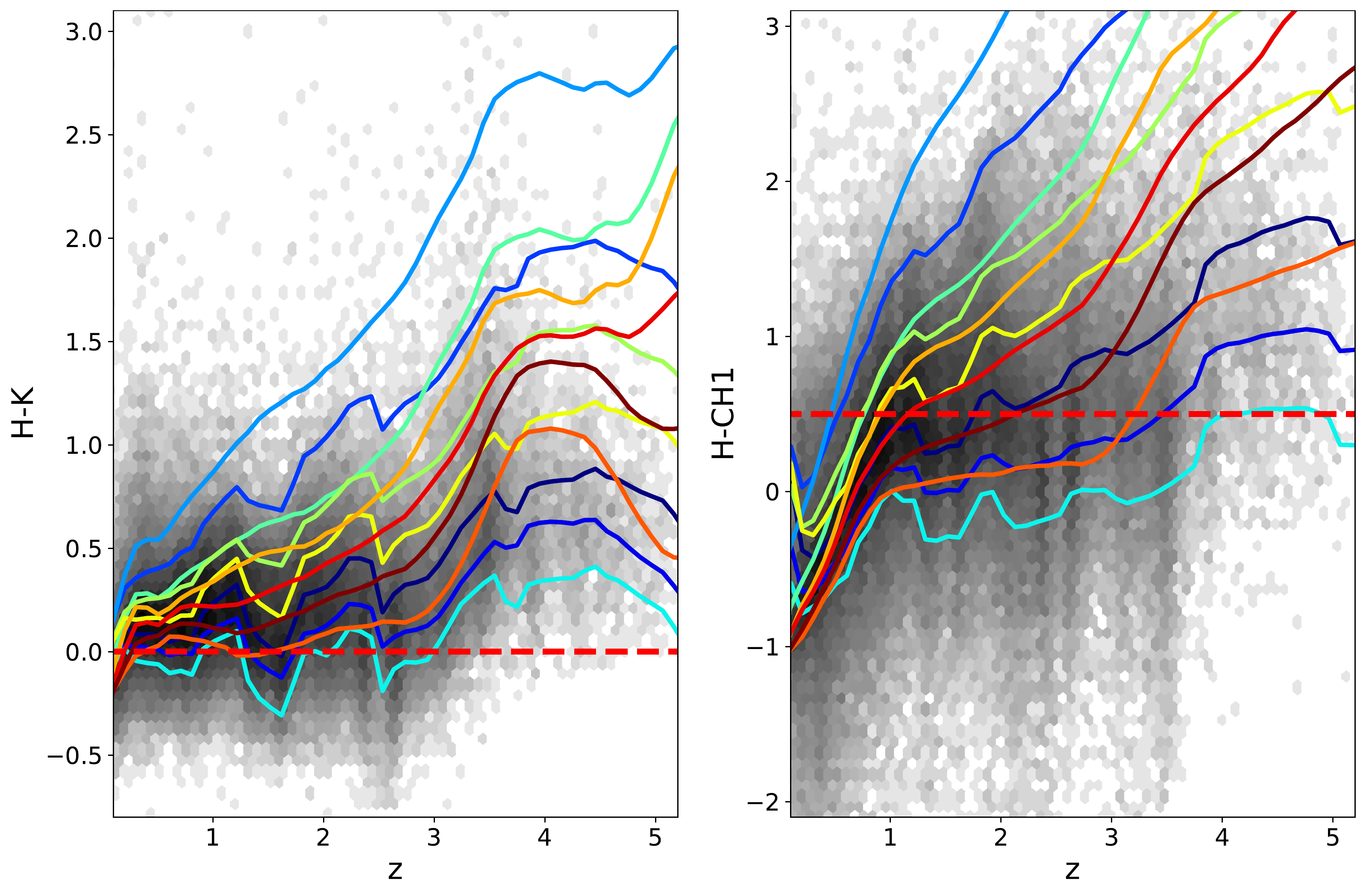}
    \caption{Distribution of $H-K$ and $H-[3.6]$ colors for the full UltraVISTA parent sample as a function of redshift. The colored curves in each panel show the redshift evolution of the colors of the EAZY templates used to determine redshifts for the base UltraVISTA DR3 catalog. The templates cover the range of observed colors for the bulk of the population, but fail to capture the most extreme red colors observed. The red horizontal dashed lines show the values chosen to select our initial sample to be remodeled with FAST++. \label{fig1}}
\end{figure*}

\section{Sample Selection and Redshifts}

Our analysis begins with the UltraVISTA DR3 catalog, which provides redshifts calculated with EAZY \citep{brammer08} and stellar population parameters with FAST \citep{kriek09}. For reasons outlined below, we have decided to re-derive both redshifts and stellar population parameters for our sample. To this end we perform an initial fit of the UV-MIR photometry with FAST++ \citep{schreiber18} to obtain redshifts and stellar population parameter estimates with which to perform our selection. Finally, we fit the UV-FIR SEDs of this final sample with the high-z extension of MAGPHYS \citep{dacunha08,dacunha15} to obtain more robust estimates of the star-formation rates and dust properties.

\subsection{Initial Catalog}

There is a well known degeneracy in reddening due to redshift, stellar age, and dust attenuation. Any successful photometric redshift estimation must therefore allow for the full range of these effects. \citet{marchesini10} were the first to show using the photometric redshift code EAZY \citep{brammer08} that introducing a template that is both old \textit{and} dusty moves a significant number of sources that would be incorrectly identified as massive galaxies at $z>3$ to $z \sim 2-3$. The effects introduced by including such a template were further explored in \citet{muzzin13b} and \citet{marchesini14}. With sufficient dust, the ability to detect the Lyman break is reduced, and the Balmer and 4000 \r{A} breaks become broadened, resulting in both less accurate and less precise photometric redshifts. These findings illustrate the importance of including the widest possible range in stellar population parameters/templates when calculating photometric redshifts, or equivalently, that the adopted template set spans all the possible ranges in the color-color spaces as real galaxies.  

We choose to re-derive the redshifts for our UltraVISTA sample rather than use the default EAZY redshifts from the catalog in order to ensure that we include photometric solutions for the most extreme cases of dust obscuration. For an initial sample to perform our revised redshift modeling, we select sources from the UltraVISTA DR3 v3.3 catalogs with $K_S<25$ and colors $H-K>0$ and $H-[3.6]>0.5$. These color cuts are intentionally relaxed compared to those used to select extremely red objects (EROs) and similar objects in the literature in order to be as comprehensive as possible in our selection of dusty objects. These cuts will only exclude blue and obviously unobscured sources from our sample. Additionally, a signal to noise ratio cut at 5 is applied to the $H-$ and $[3.6]-$ bands in order to ensure a reliable color selection (the magnitude cut ensures a signal to noise ratio greater than five for the $K_S$-band). Figure 1 shows the distribution of $H-K$ and $H-[3.6]$ colors as a function of redshift for the full UltraVISTA parent sample ($\sim 60,000$ sources) with the color cuts indicated as red horizontal dashed lines. Redshifts in this figure are the standard v3.3 version calculated with EAZY. Additionally, each panel shows the color evolution with redshift for the EAZY template set used to derive the redshifts. The twelve templates shown are meant to span a wide range of galaxy types such that the SED of an observed source can be fit through linear combination of the templates. The color tracks show that EAZY requires the reddest templates in order to match the observed colors for the most extreme sources, and that even for these objects, the template set is pushed to its limit. There is an inherent trade-off in constructing a template set that can act as a basis for a diverse sample of galaxies, simultaneously covering the full range of observed SEDs and limiting the number of templates to prevent over-fitting. It is expected that edge cases may not be fit well by the templates. For this reason, we choose to re-derive redshifts for the sample meeting the above color cuts allowing for a wider range in stellar population model parameters.

\subsection{FAST++ Modeling}

The 30,205 sources which meet both color selections are modeled with FAST++ \citep{kriek09, schreiber18} with a Calzetti dust attenuation law \citep{calzetti00}, a \citet{chabrier03} IMF, delayed exponentially declining star formation history, and allowing $0 \le $A$_V \le 6$ in order to redetermine their redshifts. FAST++ utilizes the full list of 49 photometric bands spanning the UV to \textit{Spitzer} IRAC 8$\micron$ in the UltraVISTA DR3 catalog. In addition, FAST++ also allows the user to provide an IR luminosity to help break the age-dust degeneracy inherent when fitting UV-NIR data. Providing a luminosity requires knowing the redshift beforehand. Our methodology is to fit the UV-8$\micron$ photometry to determine the redshift, then perform a second fit with the redshift fixed while providing an IR luminosity obtained by scaling the observed MIPS 24$\micron$ flux using the calibration of \mbox{\citet{dale02}}. This obviously does not provide an optimum estimate of LIR, but does provide a way to help break the age-dust degeneracy. 

Through comparison with the original EAZY redshifts we decided upon the following approach. Given the default FAST++ setup with these modeling assumptions, we find that $\sim 2\%$ of sources have failed fits. These failed fits either return models which fail to match the observed photometry or have unrealistic physical parameters. For these sources, we rerun FAST++ without a template error function to obtain a redshift. With this redshift, we run FAST++ with the redshift fixed at this value and re-instituting the template error function to derive the stellar population parameters. For the majority of initially failed fits, this process provides reasonable results. Figure 2 shows the results of our comparison of $z_{\rm FAST++}$ and $z_{\rm EAZY}$ with coloring according to the A$_V$ derived by FAST++. The black line indicates the 1:1 relation, and the dashed lines are at factors of 0.5 and 1.5 below and above. In general there is good agreement. Although there are always systematic differences between photometric redshift codes, the agreement for the bulk of the sample shows our FAST++ redshifts to be reliable. We remove outliers lying outside the dashed lines as we find they are almost universally failed fits. Of particular interest to this work are sources for which FAST++ prefers a solution with high levels of obscuration at lower redshift, which can be prominently seen in the $z=1-3$ range. These sources would potentially be missed with modeling that does not allow sufficient levels of dust obscuration.

\begin{figure}
\centering
\includegraphics[width=\columnwidth]{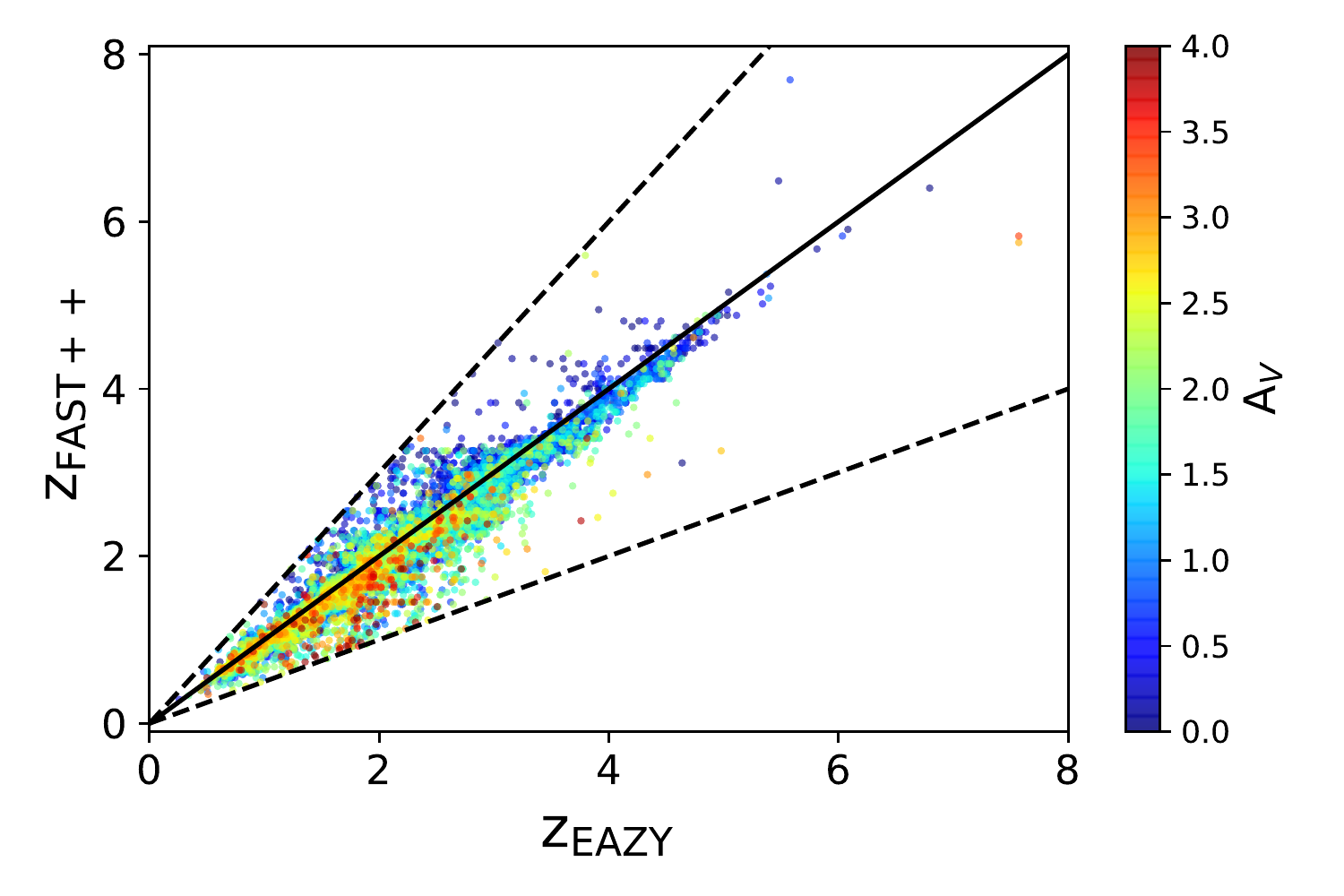}
\caption[width=\columnwidth]{Comparison of redshifts derived by EAZY and FAST++. Points are colored according to $A_V$ corresponding to $z_{\rm FAST++}$. The one-to-one line is plotted in solid black. The dashed black lines indicate difference of factors of 0.5 and 1.5. Outliers beyond this region are excluded from the analysis. \label{fig4}}
\end{figure}

\begin{figure*}
\includegraphics[width=\textwidth]{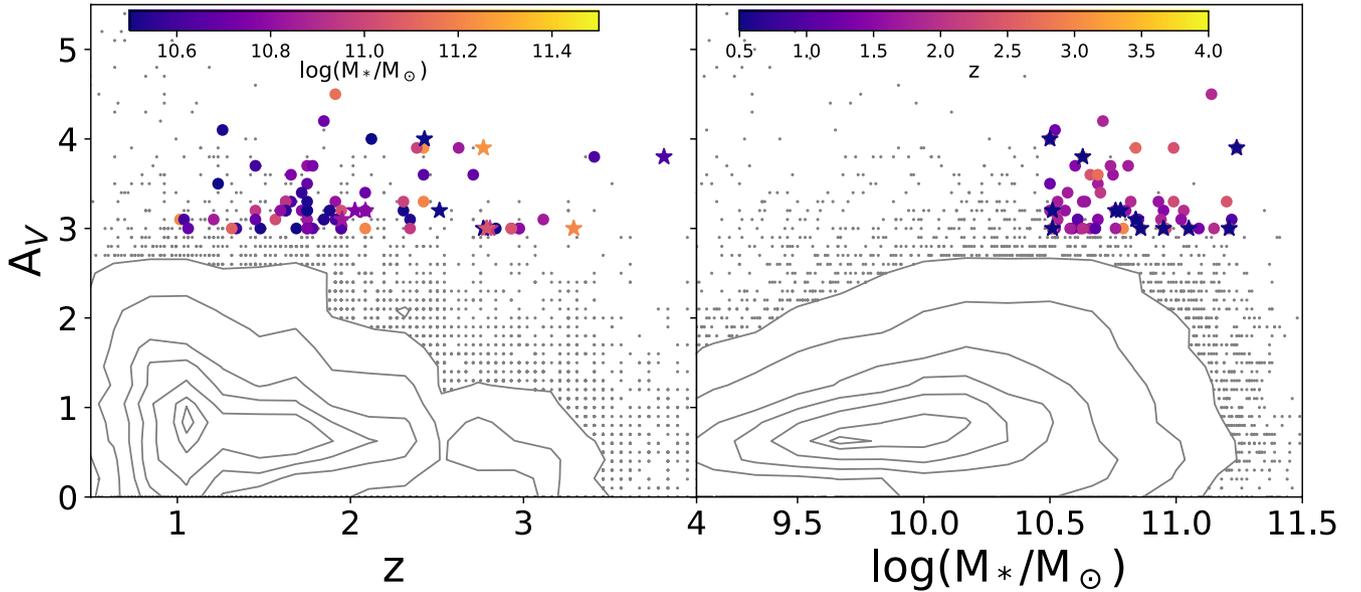}
\caption{Relations between A$_V$, $z$, and M$_*$ for sources we remodeled with FAST++. Contours show the density of sources for the full sample at the 5, 15, 35, 50, 65, 85, and 95 \% levels while gray points show sources outside the 95$^{th}$ percentile range. Colored points show the final sample and are shaded by the remaining of the three parameters in each plot. Stars indicate AGN.\label{fig-avmz}}
\end{figure*}

Figure \ref{fig-avmz} shows the relations between $z$, A$_V$, and M$_*$ for the sources we model with FAST++. Contours and small gray points show the distributions for all sources we remodeled, while colored points indicate our final selection. Using the output parameters from FAST++, we select sources with $1 \le z \le 4$, A$_V \ge 3$, and log(M$_*/\msun) \ge 10.5$. The attenuation cut can be seen in Figure \mbox{\ref{fig-avmz}} to select only the most highly obscured sources from the parent sample, with the distribution sharply dropping off after this point. We found no reliable candidates beyond $z=4$, and so adopt this value as our limiting redshift. Finally, we chose a mass limit that allows us to be mass-complete in our highest redshift bin. This additional selection is a requirement to properly interpret any trends with stellar mass, including the star-forming main sequence and size-mass relation (see Sections 4.3 and 4.4). This process results in a final sample of 65 sources. Figure \ref{fig-avmz} illustrates that these sources indeed lie at the extremes of physical parameter space. These sources are matched to the \herschel  FIR data as described in Section 2. 
Table 1 lists the detection rates for the MIPS 24 $\micron$ and \herschel bands for both our final sample and sources from the UltraVISTA catalog with $K_S < 25$ in the same redshift range. The extraordinarily high detection rates for our sample compared to the parent $K_S$-band selected sample indicate that our selection method is efficient in selecting galaxies with strong dust emission based only on the UV-to-8 $\micron$ photometry. 

To test the dependence of our selection on the assumed attenuation curve, we redetermine the redshifts and physical properties of our final sample using the \citet{kriek13} attenuation curve. With the \citet{kriek13} curve, we find that 15 of the 65 sources have disagreeing redshifts using the criterion $\Delta z/(1+z) > 0.1$. When comparing the physical parameters for sources with agreeing redshifts, we find a median offset of 0.2 mag lower A$_V$ and 0.16 dex higher M$_*$. Three of the 15 sources with disagreeing redshifts remain within the sample selection criteria. Combined with the MAGPHYS modeling (below) which uses both a different dust attenuation law and star formation history, we conclude that our SED modeling technique efficiently selects ultra-dusty galaxies, but that the precise physical properties derived will obviously depend on the assumed attenuation law. We conclude that since our selection is based on these derived properties, the exact sample selection is in part driven by model choice, but that most galaxies which meet our selection are genuinely very dusty. 

At this stage we introduce a comparison sample of sources detected with \mbox{\herschel}. From the parent UltraVISTA catalog, we select sources with identical mass and redshift cuts to the ultra-dusty sample, but in place of the A$_V$ selection, we require a 3$\sigma$ detection in at least one of the five \mbox{\herschel} bands (note this also requires a MIPS $24\micron$ detection). This results in a sample of 1616 sources that are selected based on their dust emission rather than dust attenuation.

\begin{figure}
\includegraphics[width=\columnwidth]{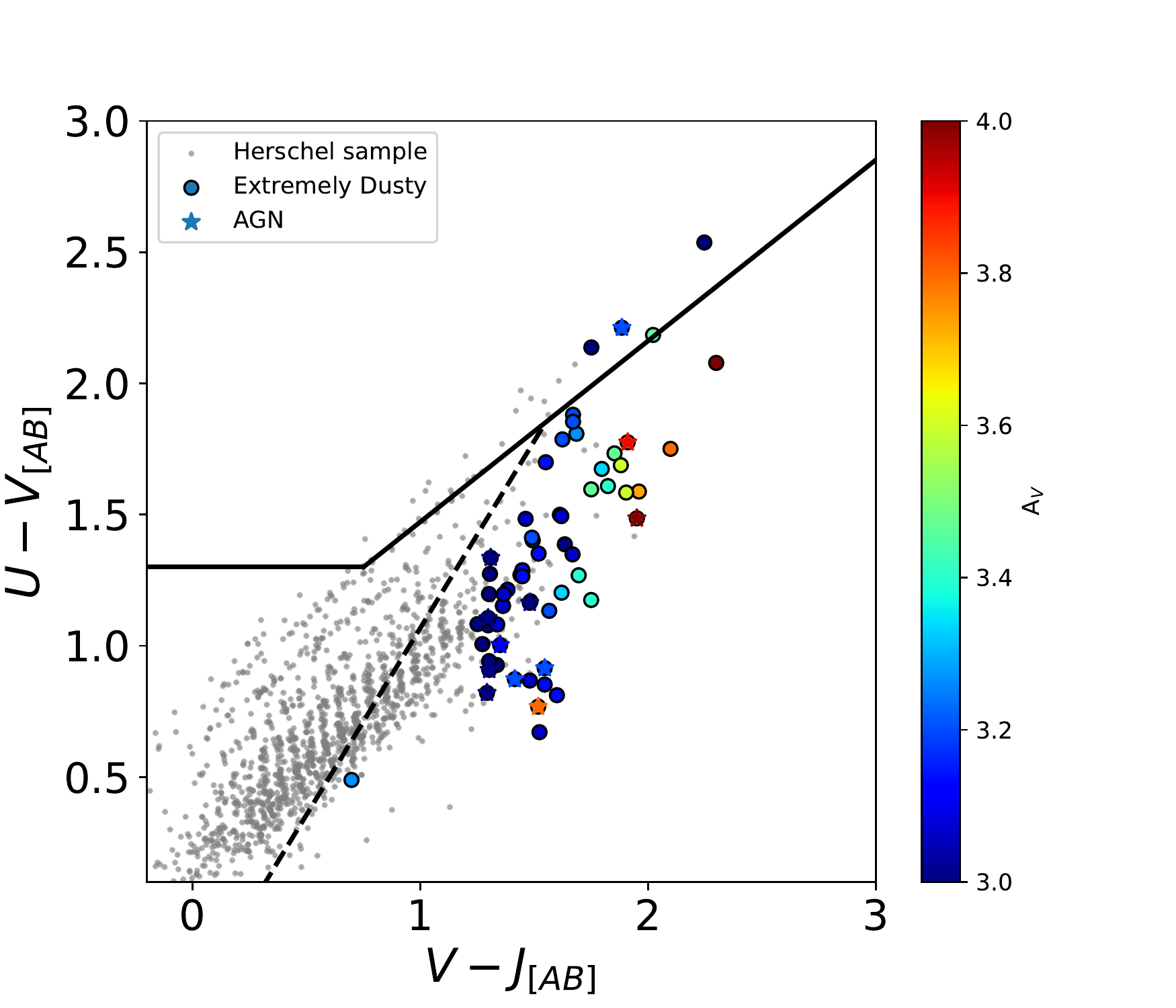}
\caption[width=\columnwidth]{Rest-frame $U-V$ and $V-J$ colors for our final sample of 65 ultra-dusty galaxies (colored) and the \herschel comparison sample (gray). The dashed line indicates the division between dusty and non-dusty star forming galaxies derived in \citet{martis16}. Points are colored by A$_V$. Stars indicate AGN. \label{fig-uvj}}
\end{figure}

As an additional check of the robustness of our selection, we examine the rest-frame UVJ colors of our final sample of sources in Figure \ref{fig-uvj}. Points are colored by their A$_V$ value. We find that all but four of our sources (94\%) fall in the dusty star-forming region of the UVJ diagram. The remaining four reside in the quiescent region, although we point out that these sources lie near the boundary. We conclude that our selection based on the FAST++ A$_V$ is generally consistent with a UVJ color selection when the attenuation reaches our chosen level of 3 mag. Comparatively, in \citet{martis19} we demonstrate that dusty galaxies with less severe attenuation A$_V \gtrsim 1$mag) occupy the quiescent and non-dusty star-forming regions of the UVJ diagram in higher numbers. We also show the UVJ colors of the \mbox{\herschel} comparison sample in gray. The \mbox{\herschel} sample occupies almost exclusively the star-forming region of the diagram, but with significantly bluer colors than the ultra-dusty sample. This suggest our sample is extreme in color and dust attenuation not only with respect to the overall galaxy population, but compared to FIR-selected galaxies as well.

\subsection{MAGPHYS Modeling}

We remodel the final sample of 65 sources using the high-z extension of MAGPHYS \citep{dacunha08, dacunha15} including the matched \herschel data in order to obtain intrinsic, unattenuated SEDs to which we compare the observed SEDs. The MAGPHYS modeling utilizes most of the UV-MIR bands as well as fully incorporating the MIPS 24$\micron$ and \textit{Herschel} measurements for 49 bands total. Sources without matched \herschel counterparts are assigned upper limits for the \herschel bands. MAGPHYS treats the upper limits by setting the flux to zero and the error to the upper limit for any nondetections. We use the photometric redshifts from FAST++ when computing the models. For the values of the physical parameters, we adopt the medians of the output distributions. See Appendix A for further description of the MAGPHYS modeling.

\begin{figure*}
\includegraphics[width=\textwidth]{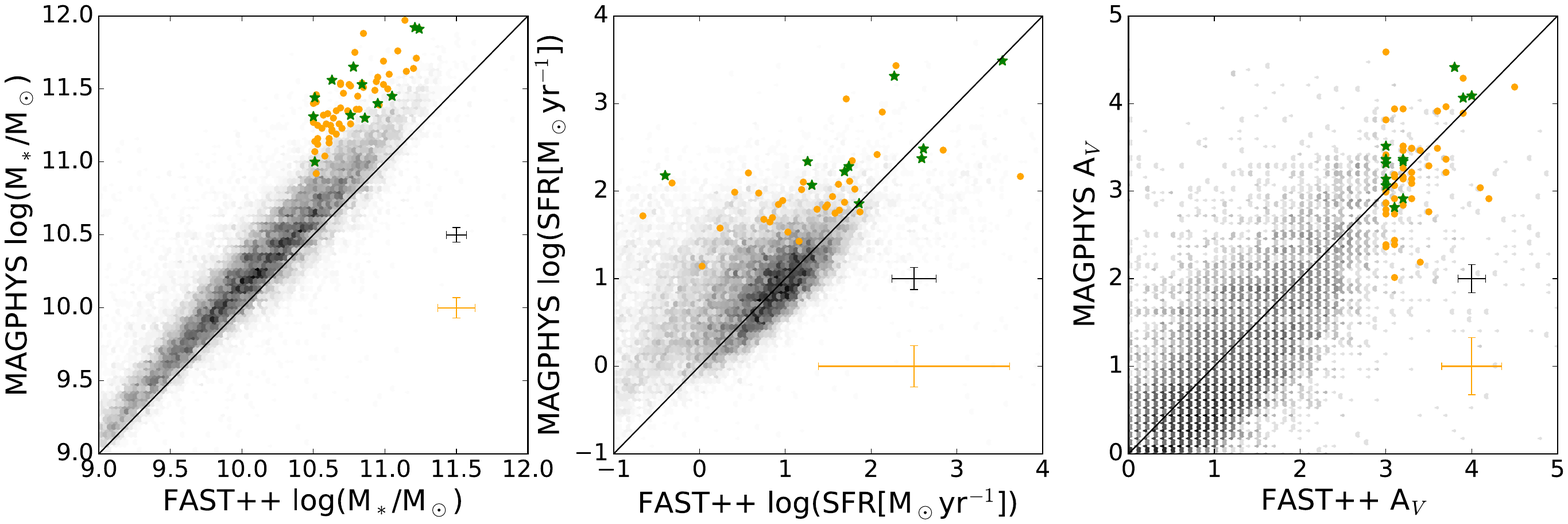}
\caption{Comparison of the physical parameters derived by FAST++ and MAGPHYS for all sources which meet our original color selection.  The full color-selected sample is shown in grayscale, and the final extreme dusty galaxy sample is shown as orange points. Green stars indicate AGN. \label{fig-fast-magphys}}
\end{figure*}

Figure \ref{fig-fast-magphys} shows the comparison of physical parameters derived by FAST++ and MAGPHYS. We observe that the stellar population parameters are comparable to the FAST++ output values, although we do note a systematic increase in the derived stellar masses of $\sim 0.2$ dex when using MAGPHYS, with the offset being more pronounced for our extreme dusty galaxies. From the above testing, we found that a differing attenuation law can cause discrepancies of this magnitude, but the degree to which the star formation history or inclusion of FIR data contribute is not clear. We note that the star formation histories used in our FAST++ modeling and MAGPHYS are qualitatively similar, allowing a rising segment at early times, and exponential decline at late times. The SFR from FAST++ and MAGPHYS are in generally good agreement, although it appears there is a tail of sources with SFR from MAGPHYS being larger, potentially caused by the inclusion of the IR information and better accounting of dust-obscured star formation. A$_V$ from FAST++ and MAGPHYS are also surprisingly well correlated, albeit with a large scatter of 0.35 mag. In general, A$_V$ and SFR are notoriously the most uncertainly derived properties from SED modeling \citep[e.g.,][]{muzzin09}.

\begin{figure}
\includegraphics[width=\columnwidth]{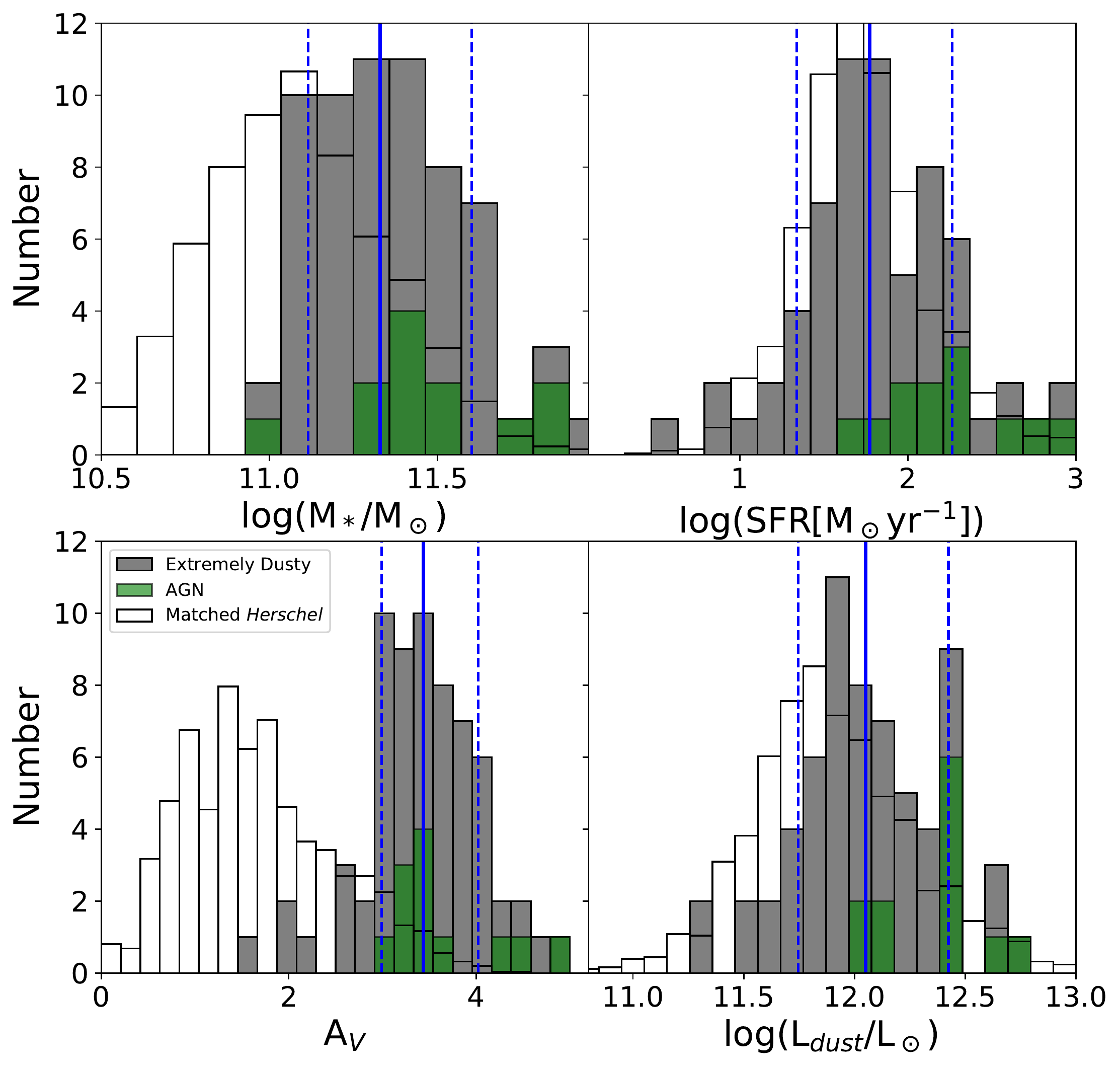}
\caption{Distributions of key physical parameters for the final sample of extreme dusty galaxies derived from the MAGPHYS UV-to-FIR SED modeling (gray). Medians of the distributions are shown as blue lines while dashed blue lines indicate the 16$^{th}$ and 84$^{th}$ percentiles. Green histograms show the distributions for sources identified as AGN. Black outlined histograms show distributions for the \mbox{\herschel} comparison sample scaled to match the number of sources in the extremely dusty sample.\label{fig-magphys-hist}}
\end{figure}

The distributions of key physical parameters of the MAGPHYS fits are shown in Figure \ref{fig-magphys-hist}. Since MAGPHYS assumes a different treatment for the dust \citep{charlot00}, the fact that only eight of the 65 sources in the final sample fall below $A_V=3$ mag  reinforces the robustness of our selection of heavily dust-obscured objects. All 65 objects are classified as (U)LIRGs according to the output L$_{dust}$ from MAGPHYS, further supporting this conclusion. The SFRs for this sample span a wide range with a median of 95 $\msun$ yr$^{-1}$. We also show the physical parameter distributions derived from MAGPHYS for the \mbox{\herschel} comparison sample, scaled to match the number of objects in the ultra-dusty sample. Several interesting points immediately emerge. The \mbox{\herschel} sample tends to slightly lower stellar masses ($\sim 0.2$ dex), in broad agreement with observed correlations between stellar mass and dust attenuation \mbox{\citep[e.g.][]{whitaker10}}. Interestingly, the comparison is well matched in star-formation rate and dust luminosity, meaning that the extreme optical attenuation levels do not significantly bias the sample to higher levels of dust emission than typical \mbox{\herschel} samples. Finally, we find that our ultra-dusty sample exhibits typical A$_V$ close to two full magnitudes higher than the \mbox{\herschel} sample. We see that the high-$A_V$ tail of the \mbox{\herschel} sample reaches these attenuation levels, but only in small numbers. Such a radical difference suggests that these objects either differ from other samples of dusty star-forming galaxies in terms of their dust properties or represent the high obscuration tail of currently known populations of dusty galaxies. This may be caused by different dust composition, geometry, mass, or any combination of the above. Some possible explanations are explored below. For more discussion of how the \textit{Herschel} comparison sample relates to our selection, please see Appendix D.

\begin{table}
    \centering
    \begin{tabular}{c|c|c}
    Band     &  Detection Fraction (\%) & Detection Fraction (\%) \\
        & Massive and Dusty & Parent Sample \\
    \hline
    MIPS 24 $\micron$     & 86 & 10 \\
    PACS 100 $\micron$     & 8 & $< 1$ \\
    PACS 160 $\micron$     & 9 & $< 1$ \\
    SPIRE 250 $\micron$     & 60 & 3 \\
    SPIRE 350 $\micron$     & 45 & 2 \\
    SPIRE 500 $\micron$     & 31 & 1 \\
    \end{tabular}
    \caption{Fraction of sources in our final sample (see text) detected at 24 $\micron$ and in each \textit{Herschel} band compared to a sample matched in $K_S$ magnitude and redshift.}
    \label{Table 1}
\end{table}

\section{Results and Discussion}
\subsection{Physical Properties and Median SEDs}
\begin{figure}
\includegraphics[width=\columnwidth]{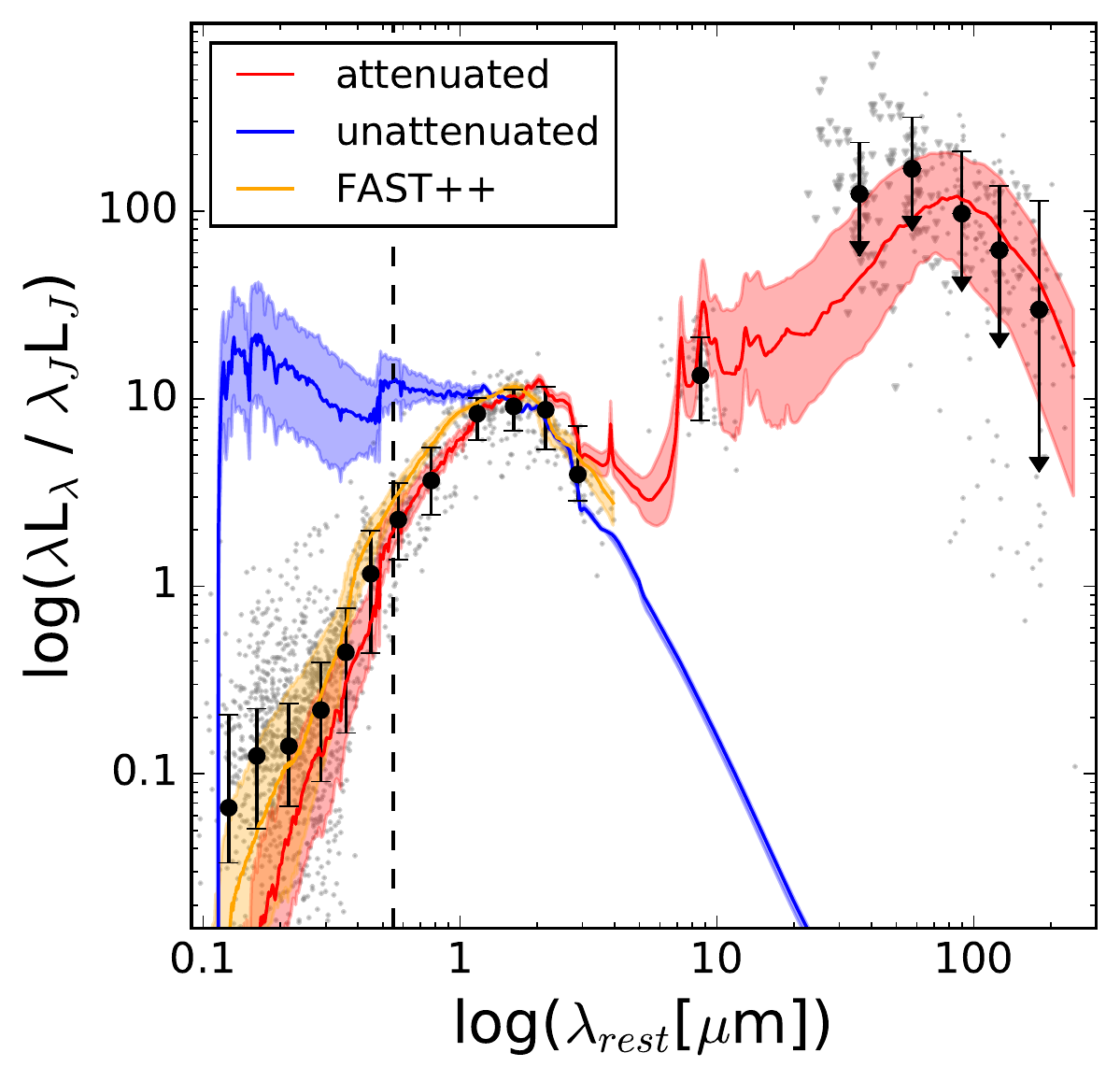}
\caption[width=\columnwidth]{Median of the best fit rest-frame SEDs output by MAGPHYS for the final sample. Red shows the attenuated SED whereas blue shows unattenuated. The shaded regions indicate the range from the $15^{th}$ to $85^{th}$ percentiles. The SEDs are normalized to the rest-frame $J$-band, approximating a normalization by stellar mass. Individual photometric observations corresponding to the attenuated SED are shown as small gray points. For the \herschel bands, 3$\sigma$ upper limits are indicated by triangles in the case of non-detections. The median photometry as well as its $15^{th}$ to $85^{th}$ percentiles are plotted as large black points with errorbars. The downward arrows for the \herschel points indicated that the 3$\sigma$ limits are used in the calculation of the median. For comparison, the median best fit FAST++ model with $15^{th}$ to $85^{th}$ percentiles is shown in orange. \label{fig-sed}}
\end{figure}

A selection at A$_V \ge 3$ implies that fewer than $\sim 5\%$ of optical photons are able to escape their region of production. Such extreme attenuation levels are usually only observed within the central regions of starbursts or in some SMGs \citep[e.g.,][]{casey17, calabro18, schreiber18}, but the presented analysis suggests a sample of objects with similar attenuation averaged over the \textit{entire galaxy}. To illustrate the extreme effect of dust on these galaxies' SEDs, we show in Figure \ref{fig-sed} the attenuated (red) and unattenuated (blue) SEDs for our sample derived from the MAGPHYS modeling. The solid curves show the median SEDs for all sources in the sample shifted to the rest-frame. The shaded regions indicate the range from the $15^{th}$ to $85^{th}$ percentiles. The SEDs are normalized to the rest-frame $J$-band, effectively normalizing by stellar mass. Individual photometric observations are shown as small gray points. For the \herschel bands, 3$\sigma$ upper limits are indicated by triangles in the case of non-detections. The median photometry as well as its $15^{th}$ to $85^{th}$ percentiles are plotted as large black points with errorbars. The downward arrows for the \herschel points indicate that the 3$\sigma$ limits are used in the calculation of the median. By visualizing the SED in terms of luminosity, one can see that almost all of the energy output for these objects occurs at IR wavelengths. In agreement with the FAST++ A$_V$ selection, the attenuated and unattenuated SEDs' luminosity differ by a factor of $\sim 100$ in the optical region, and even more in the UV (the normalization in Figure 9 somewhat reduces the appearance of this). We also show for comparison the median best fit FAST++ SED and $15^{th}$ to $85{th}$ percentiles in orange. In the overlapping region, the models agree well, with  FAST++ preferring somewhat larger luminosities in the FUV. 

The physical properties of our sample do not appear to align with typical $\herschel$-selected or SMG galaxy samples. Typical SMGs have A$_V \sim 2$ \citep{hainline11, simpson14, dacunha15}. Whereas there are examples of high levels of obscuration similar to those we observe in some of these samples \citep{ma15, ma19}, they also typically have SFRs on the order of hundreds of solar masses per year, which few of our sources reach. \citet{casey17} present NIR spectroscopy of a sample of 450 and 850 $\micron$-selected sources with a median stellar mass of $4.9 \times 10^{10}$ $\msun$, SFR of 160 $\msun$ yr$^{-1}$ and A$_V$ of 5, which much more closely align with the derived properties of our sample. Importantly, they find that 11/35 of their spectroscopically confirmed sources have incorrect or missing photometric redshifts, stressing the need for special care to be taken in determining photometric redshifts of such heavily obscured sources. They argue that the spatial disconnect of obscured and unobscured regions poses the largest problem in this regard. Addressing this issue for our sample would require high resolution imaging of the thermal dust continuum to identify the physical regions responsible for the observed IR emission.

\citet{simpson14} and \citet{elbaz18} also argue strongly that the spatial configuration of stars and dust presents a risk to interpreting panchromatic SED modeling. In their analysis of a sample of ALMA-selected SMGs, the former find that the stellar light extends to radii 3-4x further out than the dust emission of the central starburst, meaning that energy balance arguments will no longer apply, and making estimates of physical properties unreliable. In fact, when they estimate dust extinction using a hydrogen column density inferred from their gas mass estimates, they find an average A$_V=540$ mag. Clearly this disagrees strongly with any results obtained through SED modeling. Resolution of this issue at high redshift will require detailed, spatially resolved studies of the stars and dust, which will likely require observations with the \textit{James Webb Space Telescope} (\jwst).

\subsection{AGN Contribution}
We wish to note the presence of active galactic nuclei (AGN) for two reasons. First, an AGN can contribute significantly to the SED of a galaxy, potentially leading to biased estimates of stellar and dust properties. Second, we are intrinsically interested in the prevalence of AGN within this population. We identify AGN within our final sample using the same criteria as \citet{martis19}. Briefly, objects are flagged as AGN based on an IRAC color selection, radio excess, and X-ray emission. In all, 12 out of 65 (18\%) meet at least one of the selection criteria. Of these 12, eight meet the IRAC color selection, seven meet the radio selection, and two are selected via their X-ray emission. Given their extreme obscuration levels, it is not surprising that the minority of the identified AGN (2/12) are X-ray AGN. Two of the sources identified as AGN have SFRs in excess of 1000 $\msun yr^{-1}$ according to both the FAST and MAGPHYS modeling, indicating that at least some sources may have their SFRs overestimated due to an AGN contribution. In all figures, AGN are identified as different symbols/colors.

\subsection{Relation to the Star-forming Main Sequence}
Previous work has shown that both dust mass and attenuation correlate with SFR \citep[e.g.][but see \citet{vandergiessen22}]{dacunha10a,nagaraj21,nagaraj22} Given the significant amount of dust attenuation in this sample, it would be reasonable to expect high SFRs. Additionally given we expect some correlation between dust attenuation and emission, highly attenuated sources would also have significant levels of dust emission, which can arise from star formation, AGN, or older stellar populations. If star formation is the primary driver of high levels of dust emission, then the source may lie above the star-forming main sequence.

\begin{figure*}
\includegraphics[width=\textwidth]{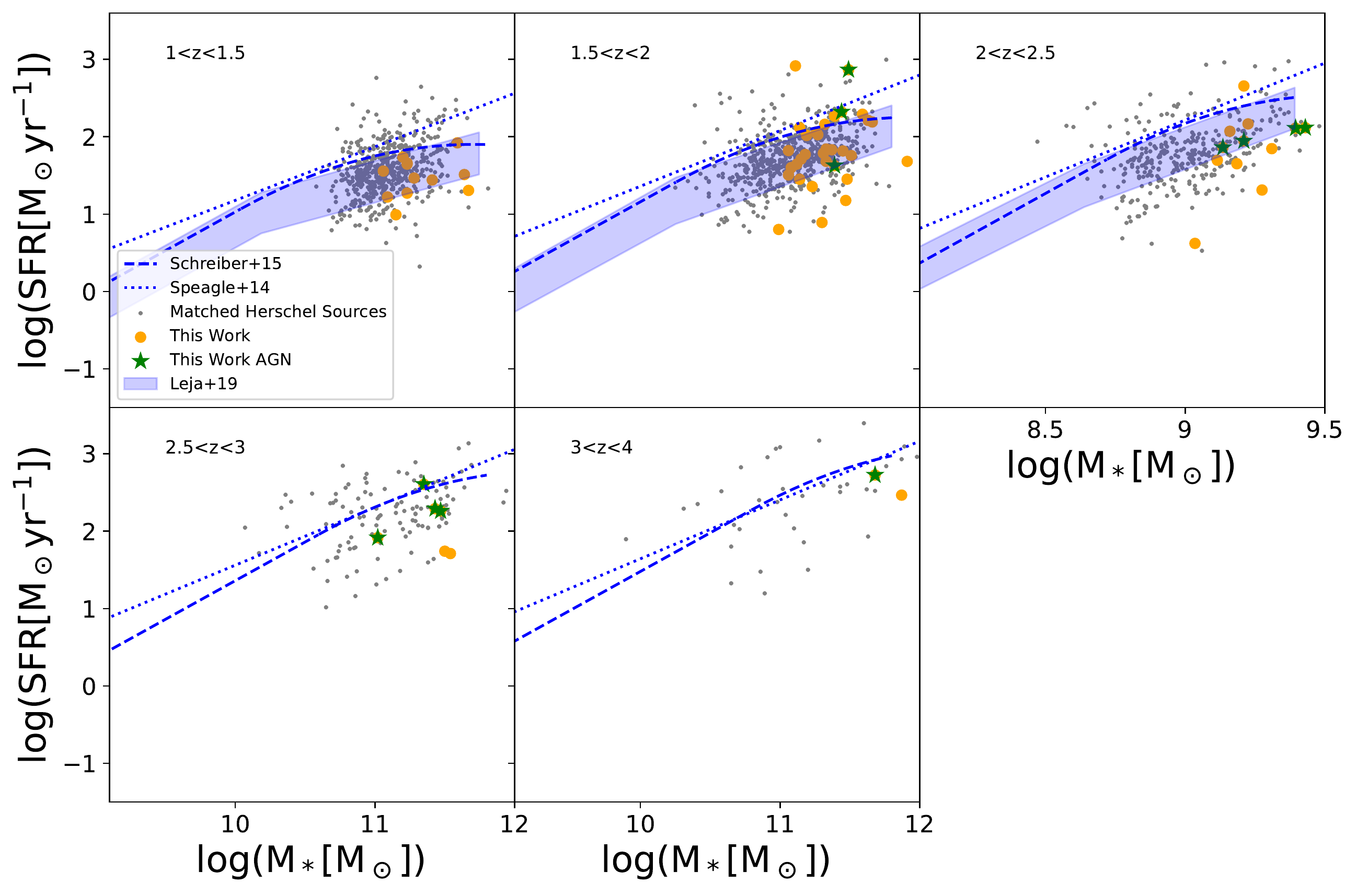}
\caption{Comparison of the SFR-M$_*$ relation for our final sample (orange points) with several relations from the literature \citep{speagle14, schreiber18, leja19}. Green stars indicate AGN and gray points show the \mbox{\herschel} comparison sample. Each panel shows a different redshift bin as indicated. \label{fig-main-sequence}}
\end{figure*}

Figure \ref{fig-main-sequence} shows the relation between SFR and M$_*$ for our final sample as well as the comparison \herschel sample along with several recent determinations of the star-forming main sequence from the literature. The SFR and M$_*$ shown are those derived from MAGPHYS. We find that the majority of our sources are consistent with lying on the main sequence and that they are not separated in parameter space from the \herschel comparison sample. This may be a somewhat surprising result, since starbursts in the relevant redshift range are frequently observed to be heavily obscured and individually detected with \herschel or in the sub-millimeter \citep[e.g.,][]{schreiber15, miettinen17, elbaz18}. \citet{martis19}, however, shows that a selection in dust obscuration leads to a sample with diverse levels of star formation. This work suggests that this holds true for stricter selections in dust obscuration as well. These results are also consistent with those of \citet{penner12} who show that their sample of DOGs which have $10^{12} \lesssim L_{\rm IR} \lesssim 10^{13}$ exhibit a range in star formation activity. Only 24\% of their sample have sSFRs which indicate the dominance of a compact starburst.

Despite the modest SFRs for our sample, we still observe significant levels of IR emission. The two component dust model utilized by MAGPHYS allows slightly older stellar populations to contribute to IR emission by heating dust in the diffuse interstellar medium rather than dust in birth clouds around star-forming regions. Recent work suggests that intermediate age stars can contribute significantly to MIR emission \citep{utomo14,leja19,roebuck19, martis19}. The level of dust heating by intermediate age stars will depend on the lifetime of dust grains generated by different sources. The observed correlation between SFR and dust mass \citep{dacunha10a} suggests that at least some dust is produced quickly by supernovae during an episode of star formation, but the degree to which older asymptotic giant branch stars can continue to produce dust well after star formation has ended remains unclear \citep{goldman21}. Depending on the dust lifetime, it may be possible that some of our sources are post-starburst galaxies which have begun to decline in SFR before the large amount of dust associated with the burst has been destroyed. Given the high levels of obscuration, and hence the relatively featureless SEDs, measuring precise star formation histories for these objects to test this hypothesis will prove extremely challenging.

\subsection{Morphology}
\begin{figure*}
\includegraphics[width=\textwidth]{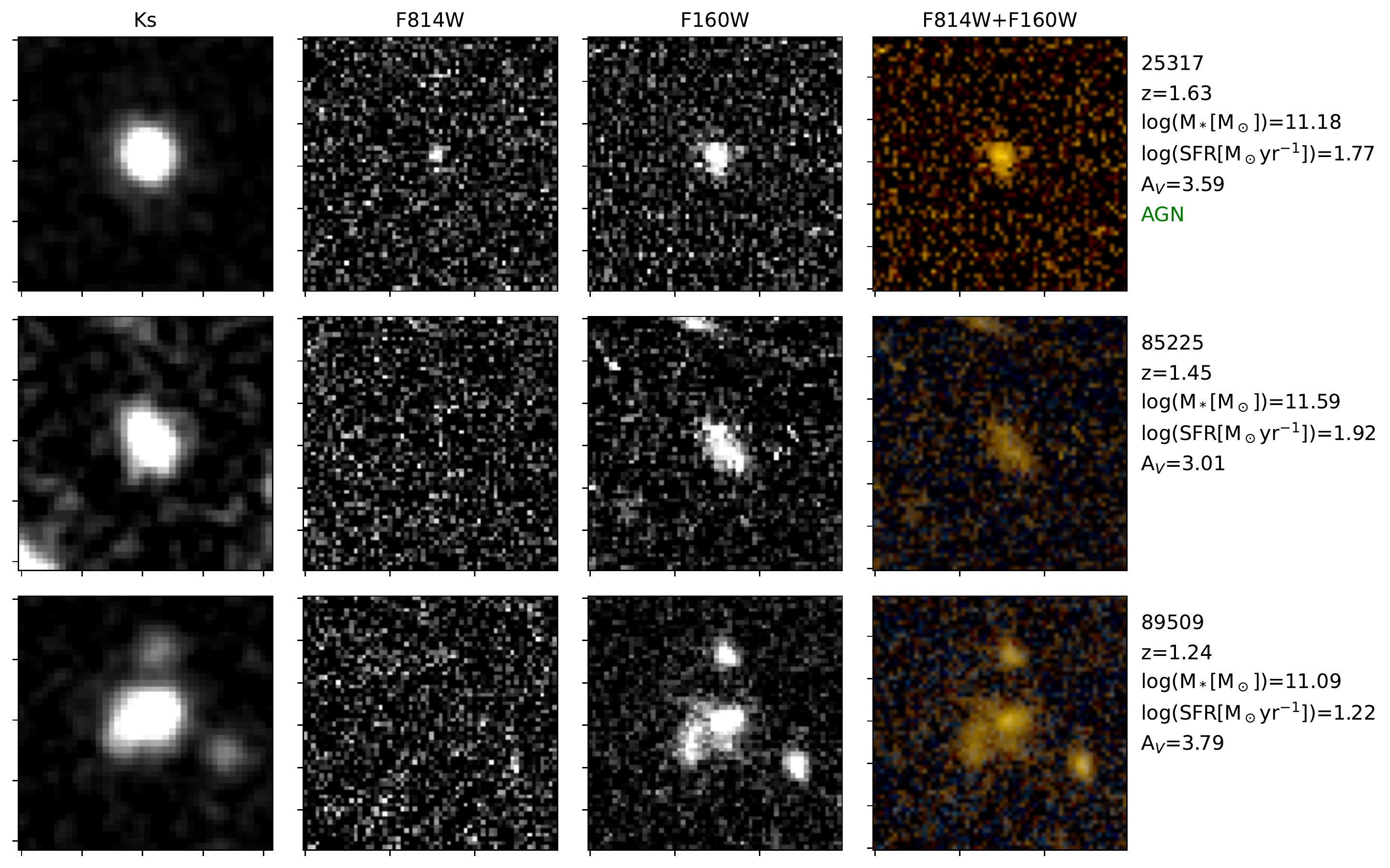}
\caption{Image stamps for three example sources in the final sample. The left column shows the UltraVISTA $K_S$-band image, center left $\hst$ $F814W$, center right $\hst$ $F160W$, and the right shows a color image constructed from the combination of the $\hst$ images. The UltraVISTA DR3 catalog number as well as the stellar mass, SFR, and A$_V$ from the MAGPHYS modeling are listed for each object. Sources identified as AGN are labeled. Images are 6" on a side. \label{fig-stamps}}
\end{figure*}

In order to better understand the physical arrangement giving rise to these conditions, we investigate the $\hst$ $F814W$ and $F160W$ images of all sources available from the COSMOS \citep{koekemoer07, massey10} and DASH \citep[Drift And SHift][]{momcheva17, mowley19} programs' coverage of the UltraVISTA survey area. Figure \ref{fig-stamps} shows three example sources. From left to right we show the UltraVISTA $K_S$, $F814W$ and $F160W$ images, as well as the UltraVISTA catalog number and MAGPHYS physical properties. We observe a range of morphologies, with some sources appearing round and compact, whereas others are more elongated suggesting a possible disk-like morphology observed edge-on. There are indications of possible mergers/interactions in progress for source IDs 89509, 159508, and 198121. 

\begin{figure}
\includegraphics[width=\columnwidth]{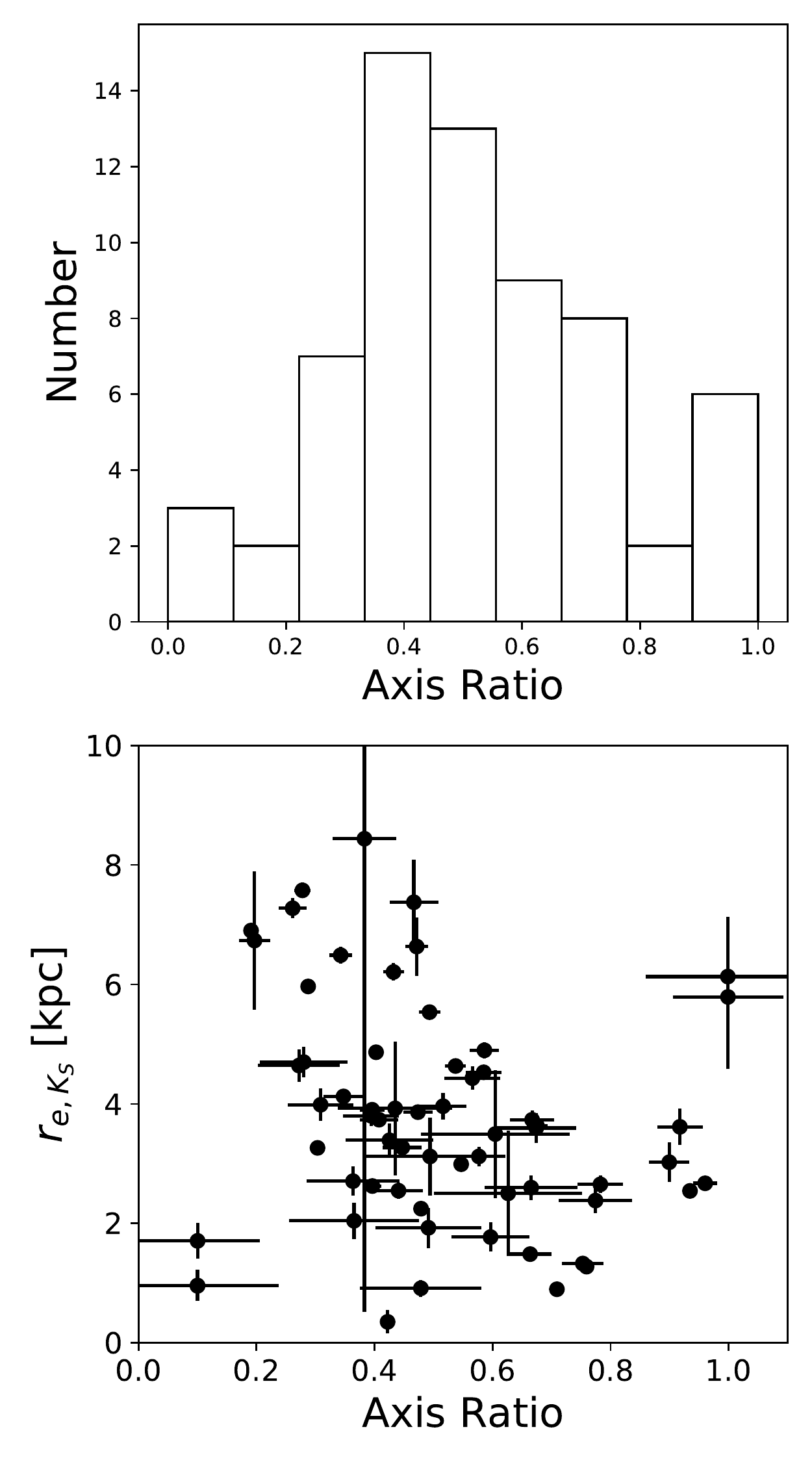}
\caption[width=\columnwidth]{Top: Axis ratios distributions for our sources derived from GALFIT from the  UltraVISTA $K_S$-band images. Bottom: Relation between effective radius and axis ratio as measured from the UltraVISTA $K_S$-band imaging from GALFIT. Errors are those reported by GALFIT. \label{fig-q-size}}
\end{figure}

To quantify these observations, we perform profile fitting of the $K_S$ and $F160W$ images with the code GALFIT \citep{peng10galfit}. Each source is modeled with a single Sersic profile except for the sources identified above as clearly decomposing into multiple sources in the the higher resolution $\hst$ imaging. In these cases we model each component with a Sersic model. Given that $\hst$ imaging is only available for 23 out of 65 of our sources, we reference the fitting results from the $K_S$-band images. Appendix C shows that despite the lower resolution, differences in the fit parameters for the quantities we discuss here do not affect our conclusions. 

The top panel of Figure \ref{fig-q-size} shows the distribution of axis ratios for our sample measured from the GALFIT modeling of the $K_S$-band images. We find examples of sources across the full range of galaxy shapes. The peak of the distribution occurs around an axis ratio of 0.4. Elongated sources with these lower axis ratios are likely to be disks observed edge-on. Previous work shows that viewing a star-forming disk edge-on results in a higher observed attenuation level \citep{wang18}, so this provides a natural explanation for the extreme attenuation level of our sample. 

This leaves open at least two possibilities for our sources with high axis ratios. They may represent the same population of disks, but viewed face-on, or they may be more compact starbursts such as the "blue nuggets" observed in this redshift range \citep{osborne20,zolotov15}. To begin to address this question we show in the bottom panel of Figure \ref{fig-q-size} the effective radius versus axis ratio for our sample. If these "blue nuggets" do form the rounder portion of our sample, we may expect an anti-correlation between size and axis ratio, such that rounder objects are smaller. If we discount the two sources with the highest axis ratios and large errors, then one may tentatively observe this anti-correlation. We further address this question in the next section. 

In a study of starburst galaxies at $0.5<z<0.9$, \citet{calabro18} find a wide range of obscuration values depending on location within the galaxy, with the starburst core reaching A$_V=2-30$ mag. Given that the majority of their sample are morphologically classified as mergers, they argue that these extreme obscuration values may serve as a useful tool for identifying mergers at higher redshift. Our available $\hst$ images do not allow us to rule out merger  classifications for the majority of our sources, but since only three show signs of neighboring sources, mergers do not appear necessary to generate the levels of obscuration we observe. 

\subsubsection{Sizes}
\begin{figure}
\includegraphics[width=\columnwidth]{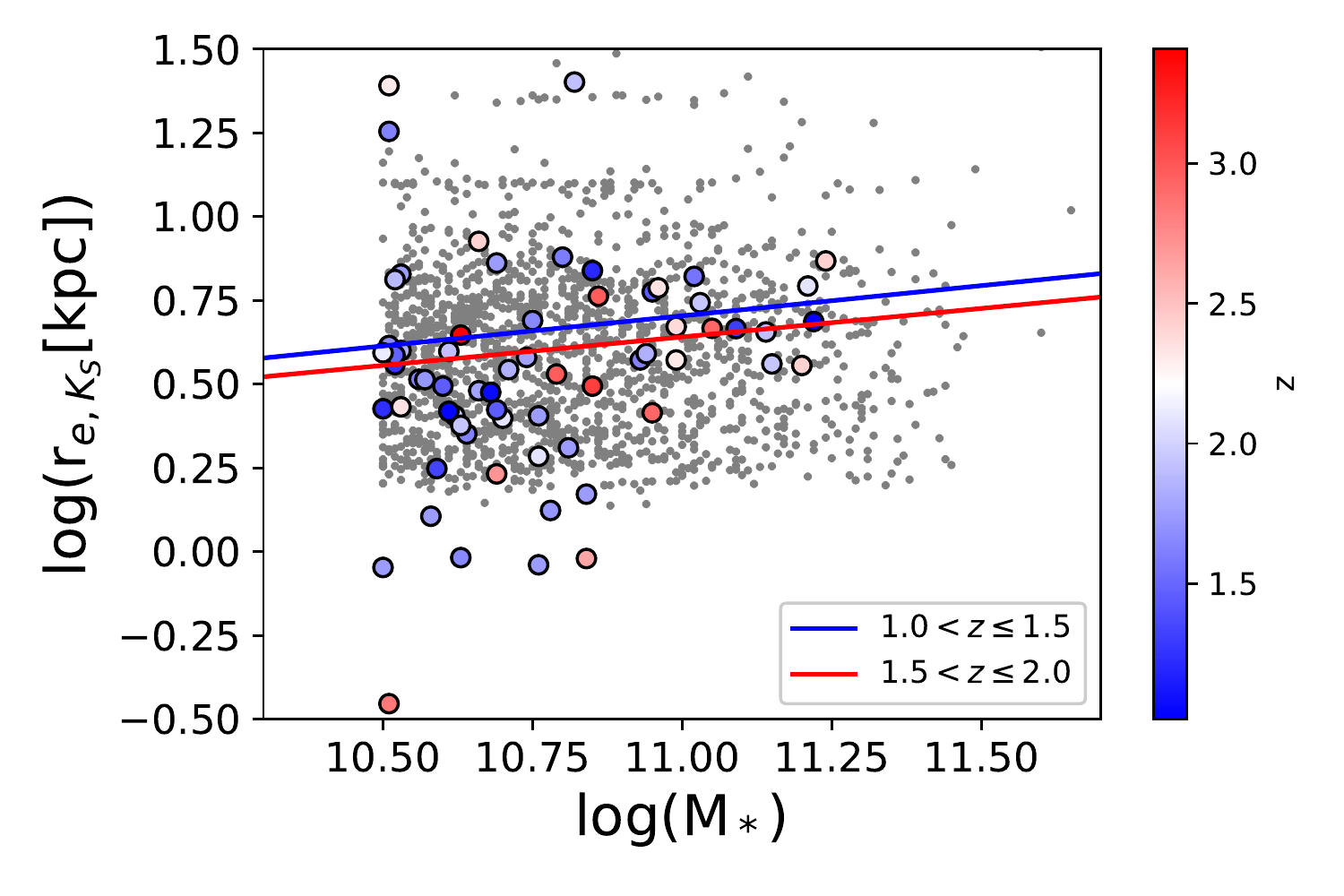}
\caption[width=\columnwidth]{Size-mass relation using K$_S$-band sizes for all 65 ultra-dusty galaxies. Points are color-coded by redshift. \mbox{\hst} $F160W$ sizes for the \mbox{\herschel} comparison sample are shown as gray points and are obtained by matching to the \mbox{\citet{cutler22}} morphological catalog of the DASH survey. The size-mass relations for star-forming galaxies in the overlapping redshift range measured by \citet{nedkova21} are shown as blue and red lines. \label{fig-size-mass}}
\end{figure}

One may expect a compact morphology to correlate with increased levels of dust attenuation due to the presence of a dust-enshrouded nuclear starburst \citep{barro16,cochrane21,dudzeviciute21}. To investigate whether this holds true for our sample, we compare the $K_S$-band sizes from our GALFIT analysis to the size-mass relation for star-forming galaxies measured by \citet{nedkova21}. Figure \ref{fig-size-mass} shows this comparison for the two overlapping redshift bins in \citet{nedkova21} with the redshift of our sources indicated by color. Sizes for the \herschel comparison sample are obtained by matching to the \mbox{\citet{cutler22}} morphological catalog of the COSMOS DASH survey \citep{mowla22}, which reports \mbox{\hst} $F160W$ sizes. The measurement wavelength differs slightly from that used for the extremely dusty sample, but is close enough to perform general comparisons. We find the bulk of our sources follow the relation for the general star-forming galaxy population with a tail extending to smaller sizes. They occupy a similar range as the \mbox{\herschel} comparison sample, with about five sources falling below the main cloud. This suggests that the extreme attenuation levels in our sample do not arise primarily from a more compact morphology across the galaxy as a whole. It is important to note, however, that we report half-light radii rather than half-mass radii. \mbox{\citet{miller22}} show that color gradients across galaxies caused by dust attenuation significantly affect the mass-size relation when measured using half-light radii. Given the extreme attenuation levels of our sample, half-mass sizes may differ significantly, but we defer this investigation to future work.

Recent work has also found evidence of compact heavily-obscured nuclear star-forming regions within more extended star-forming disks at $z \sim 2$ \citep{chen20,pantoni21}. For our redshift range, the $K_S$-band traces the rest frame optical-NIR and thus the distribution of stellar mass rather than the star-formation activity. The sizes for the majority of our sample are then consistent with the scenario of a typically-sized disk hosting a compact star-forming region which dominates the integrated star formation rate and infrared emission leading to the high levels of dust attenuation. High-resolution imaging of the dust continuum with ALMA would be necessary to test this hypothesis.

\begin{figure}
\includegraphics[width=\columnwidth]{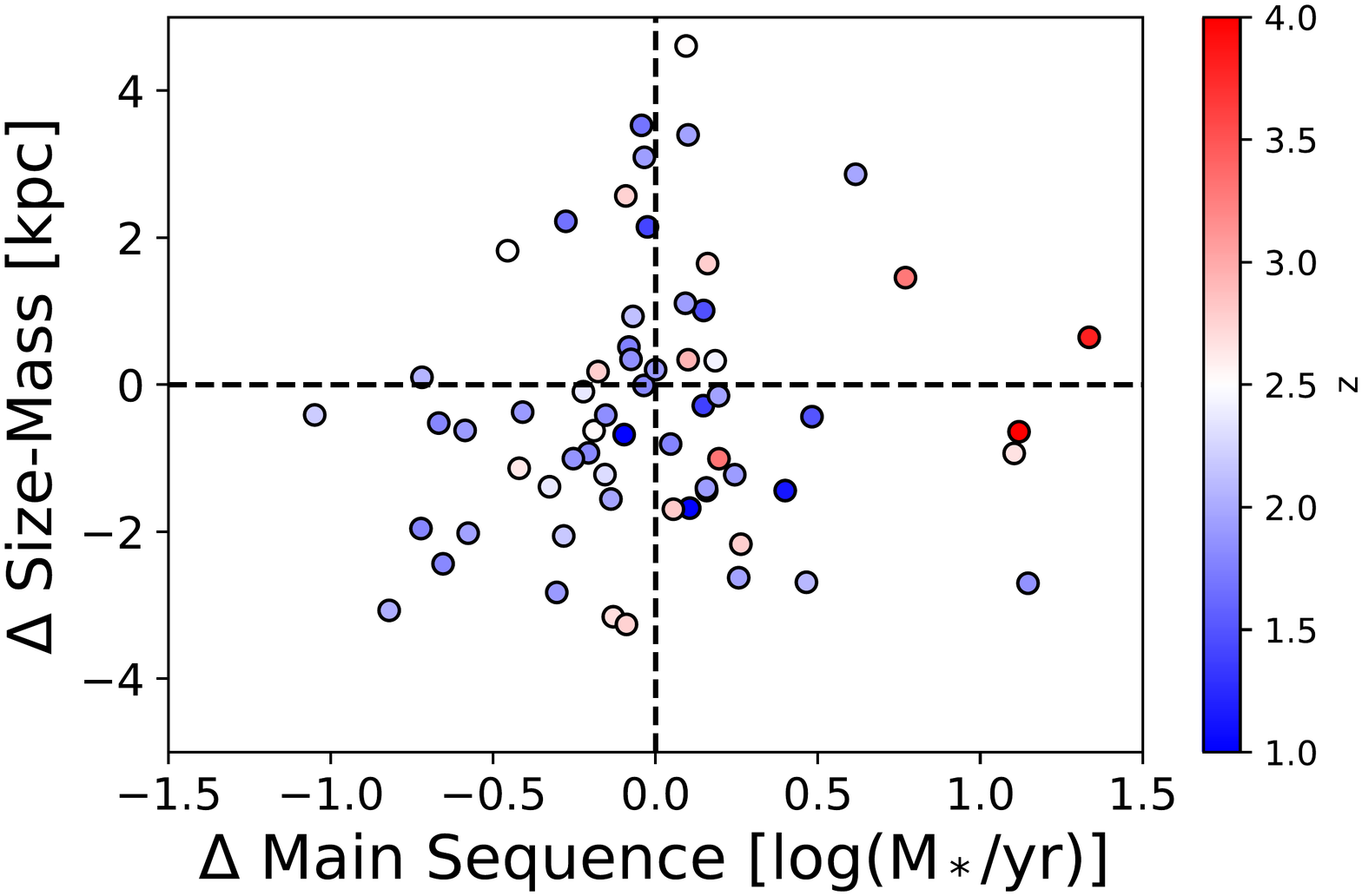}
\caption[width=\columnwidth]{Offset from the size-mass relation measured by \citet{nedkova21} versus offset from the star-forming main sequence measured by \citet{speagle14} using K$_S$-band sizes and MAGPHYS star formation rates for all 65 ultra-dusty galaxies. Points are color-coded by redshift. \label{fig-size-ms}}
\end{figure}

\mbox{\citet{suess21}} find that dusty star-forming galaxies as well as post-starbursts are compact at $z>2$, and hypothesize an evolutionary link between these populations. This is in contrast to $z<2$, where dusty star-forming galaxies are more extended. We further investigate the possible presence of compact starbursts in our sample by comparing offset from the size-mass relation to offset from the star-forming main sequence as measured by \citet{nedkova21} and \citet{speagle14} respectively in Figure \ref{fig-size-ms}. Points are colored by redshift. We observe no strong trend between the offsets, meaning that our most heavily star-forming sources are not preferentially more compact than the rest of our sample. This means that if a central starburst does dominate the SFR for our most star forming objects, it does not drive the rest-frame optical half-light sizes smaller. We reiterate, however, that we do observe a tail of sources extending to small sizes. Given the range of SFRs in combination with the high stellar masses for these objects, it may be that we are observing the aftermath rather than the ongoing process of a compaction event. If some of these sources have recently undergone a central starburst and are in the process of quenching, this could explain the low SFRs. When they are fully quenched, they may resemble the compact, quiescent galaxies, or so-called "red nuggets" \citep{damjanov09,damjanov11}, observed at the lower end of our probed redshift range.

\section{Summary and Conclusion}
We have presented a sample of dusty galaxies from the UltraVISTA DR3 photometric catalogs with extreme levels of optical extinction. Sources satisfying color cuts at $H-K>0$ and $H-[3.6]>0.5$ are modeled with the UV-to-8 $\micron$ SED modeling code FAST++ \citep{kriek09, schreiber18} allowing for very high extinction levels. For our final sample we select sources with $1 \le z \le 4$, A$_V \ge 3$, and log(M$_*/\msun) \ge 10.5$, resulting in 65 sources. We summarize our key findings below.

\begin{enumerate}
    \item Our selection based on extinction efficiently selects galaxies with strong dust emission. 86\% of our sources are detected at 24 $\micron$ and 60\% are detected with \herschel SPIRE. Thus our sample overlaps with, but is not encompassed by the population of bright FIR emitters that make up \herschel samples.
    
    \item Full UV-to-FIR SED modeling with MAGPHYS is consistent with the FAST++ selection. The median best fit MAGPHYS SEDs show that almost all of the energy for these sources is emitted at IR wavelengths.
    
    \item The extreme obscuration levels of our sample are only matched by the high-A$_V$ tail of \herschel-detected sources of similar mass and redshift. Some FIR-selected populations, including SMGs, generally exhibit higher SFRs than are observed in our sample. Relatedly, the majority of our sample appears to be consistent with lying on the main sequence of star-forming galaxies in the corresponding redshift range. 
    
    \item Limited $\hst$ imaging of sufficient depth prevents us from making strong claims regarding the morphology of these sources, but we tentatively observe a preference for low axis ratios, suggesting a disk viewed edge-on. We only observe evidence for merging/interacting galaxies in a few cases. Therefore, a merger-induced starburst does not appear necessary for generating the levels of dust obscuration we observe.  
    
    \item The $K_S$-band sizes of our sample are generally consistent with the size-mass relation for the general population of star-forming galaxies at similar redshifts as well as with \herschel galaxies matched in mass and redshift. They are therefore not on average more compact than other star-forming galaxies, as has been observed for other dusty galaxy samples.

\end{enumerate}

Our unique approach to selecting dusty galaxies provides us a sample that differs from the conventional picture of a merger-induced starbursts or AGN providing the energy to generate an IR-luminous SED. Although our sources have large L$_{\rm IR}$, most exhibit modest star formation rates. We have investigated several potential alternative causes for the extreme dust obscuration levels and substantial dust emission, including inclined observing angles, compact sizes, and dust heating by intermediate age stars. We conclude that our selection method results in a heterogeneous sample of galaxies. Many are likely moderately star-forming disks viewed edge-on, leading to enhanced obscuration of UV-optical emission for a given SFR or L$_{\rm IR}$. Others may be more compact, rounder galaxies potentially undergoing, or having recently undergone, a compaction event. A declining star formation history in this case would lead to significant IR emission from intermediate age stars and significant obscuration if the quenching timescale is shorter than the dust destruction timescale. As noted, we also select apparently merging galaxies with low frequency. Comparatively obscured sources can be found in FIR surveys, but here we provide a novel view of the less IR-luminous segment of this population. This work provides an introduction to this sub-population of dusty galaxies, but we have exhausted much of what may be learned of these sources through UV-MIR photometry.

\section{Acknowledgements}
D.M. and N.M. acknowledge the National Science Foundation under grant No. 1513473. M.S. and N.M. acknowledge Canadian Space Agency grant 18JWST-GTO1.
Based on observations made with ESO Telescopes at the La Silla or Paranal Observatories under programme ID(s) 179.A-2005(A), 179.A-2005(B), 179.A-2005(C), 179.A-2005(D), 179.A-2005(E), 179.A-2005(F), 179.A-2005(G), 179.A-2005(H), 179.A-2005(I), 179.A-2005(J), 179.A-2005(K). 
This research has made use of data from HerMES project (http://hermes.sussex.ac.uk/). HerMES is a Herschel Key Programme utilising Guaranteed Time from the SPIRE instrument team, ESAC scientists and a mission scientist.The HerMES data was accessed through the Herschel Database in Marseille (HeDaM - http://hedam.lam.fr) operated by CeSAM and hosted by the Laboratoire d'Astrophysique de Marseille.Herschel is an ESA space observatory with science instruments provided by European-led Principal Investigator consortia and with important participation from NASA.
\\
Software: python, astropy, matplotlib, numpy, MAGPHYS, EAZY, FAST++

\section*{Data Availability}
The HerMES and PEP data that support the findings of this study are publicly available. Additional data underlying this article will be shared on reasonable request to the corresponding author.

\bibliographystyle{mnras}
\bibliography{refs}

\appendix

\section{FIR Luminosity, SPIRE Detection Rates, and SFR Measurements}

\begin{figure*}
\includegraphics[width=\textwidth]{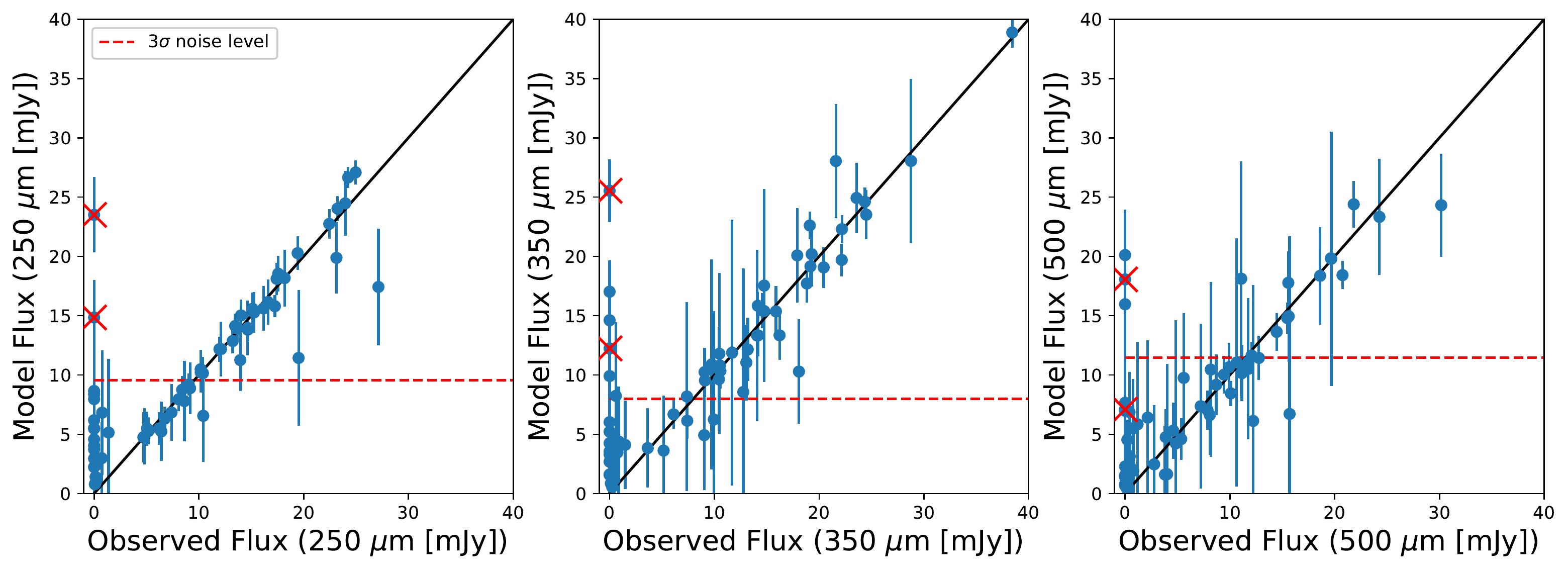}
\caption{Comparison of observed and model fluxes in the \textit{Herschel} bands for the extremely dusty sample. The red dashed line indicates the $3\sigma$ noise level in each band. The line of equality is shown in black. \label{figA1}}
\end{figure*}

\begin{figure*}
\includegraphics[width=\textwidth]{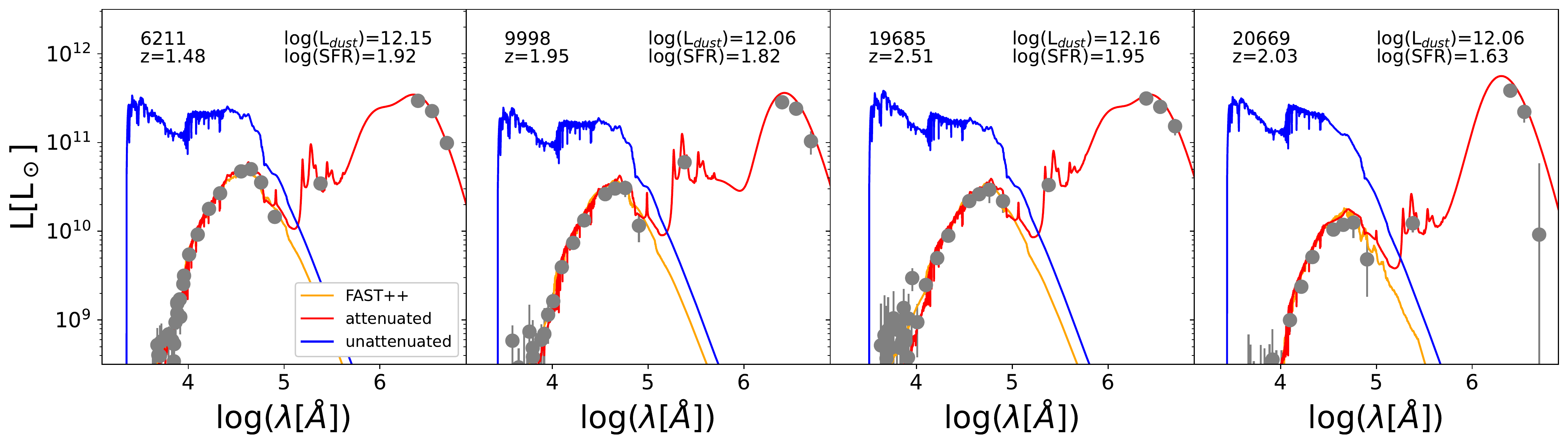}
\caption{SEDs of sources with large L$_{IR}$ and SFR $<$ 100 M$_{sun}$ yr$^{-1}$ output by MAGPHYS. Red shows the attenuated SED whereas blue shows unattenuated. Individual photometric observations corresponding to the attenuated SED are shown as gray points. For comparison, the median best fit FAST++ model is shown in orange. \label{figA2}}
\end{figure*}

In Section 4.3 we discuss the star formation properties of our sample, noting large IR luminosities (L$_{IR} > 10^{12}$ $\lsun$) and modest SFRs (SFR $<100 \msun$ yr$^{-1}$). This may be surprising to some readers, so we provide additional discussion here. Figure \mbox{\ref{figA1}} shows the comparison of observed and model fluxes for the SPIRE bands from the MAGPHYS fitting. The red dashed line shows the $3\sigma$ noise level in each band. The figure shows that when the observed flux is small MAGPHYS does not assign models with fluxes above the detection limit unless the uncertainties are consistent with a non-detection with the exception of two sources. These are marked with red x's in the plot. This first shows that despite implying considerable total L$_{IR}$, our modeling is consistent with the non-detections in SPIRE for many sources. As additional evidence, we show the SEDs of four such sources in Figure \mbox{\ref{figA2}}. The photometry is well fit by both the FAST++ and MAGPHYS models.

\begin{figure}
\includegraphics[width=\columnwidth]{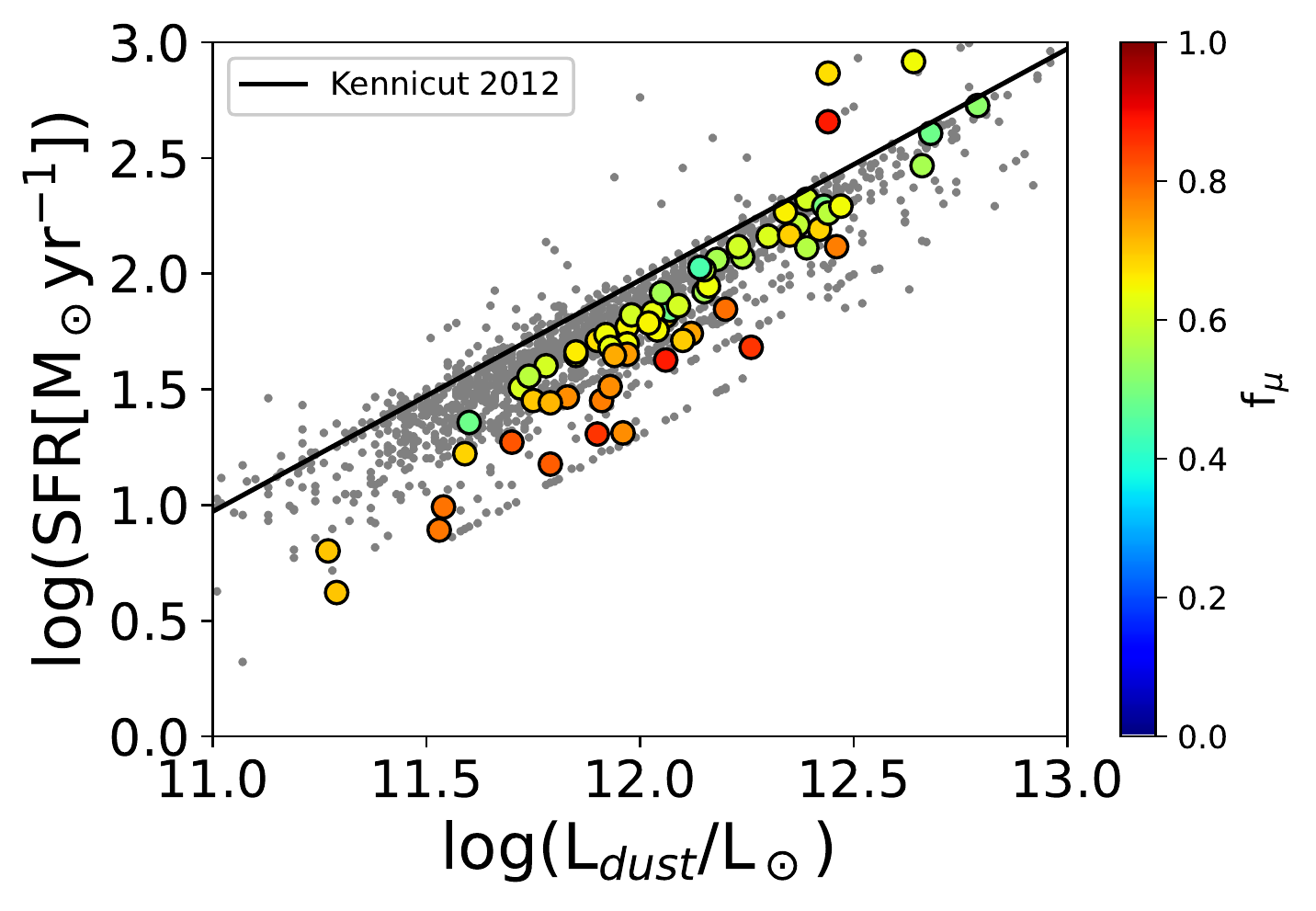}
\caption[width=\columnwidth]{Relation between total IR luminosity and SFR for the massive, dusty sample (colored) and the \textit{Herschel} comparison sample (gray). Our sample is colored by the $f_\mu$ model parameter from MAGPHYS, which denotes the fraction of the total dust luminosity attributed to the diffuse interstellar medium (rather than stellar birth clouds). The \citet{kennicutt12} relation scaled to a \citet{chabrier03} to match the MAGPHYS models is shown in black. \label{figA3}}
\end{figure}

It might still be argued that despite the satisfactory SED fits, the observed (and modeled) IR flux should still imply SFRs higher than those reported by MAGPHYS. As a final check, in Figure \ref{figA3} we examine the relation between total IR luminosity and SFR for our massive, dusty sample (colored) and the \textit{Herschel} comparison sample (gray). The two component dust model of \citet{charlot00} used in our MAGPHYS modeling allows the IR emission from stellar birth clouds, heated by active star formation, and the diffuse ISM, heated by both young and intermediate-age stars, to both be accounted for separately. Our sample is colored by the $f_\mu$ model parameter from MAGPHYS, which denotes the fraction of the total dust luminosity attributed to the diffuse interstellar medium. The \citet{kennicutt12} relation (scaled to a \citet{chabrier03} IMF to match the MAGPHYS models) is shown in black for reference. There is a clear trend showing that sources which fall further below the Kennicutt relation have more of their IR luminosity attributed to the diffuse ISM and older stars. This behavior allows a high  L$_{IR}$ in combination with unremarkable SFRs. This is fully consistent with the findings of \citet{martis19} Of course the precise timescale on which older stars contribute significantly to dust heating depends on the assumed star formation history. Spectroscopic analysis to verify the stellar ages would provide a stronger constraint, but will prove challenging for such sources given the large dust content.

\section{Alternate Star-forming Main Sequence Comparison}
\begin{figure*}[hb]
\includegraphics[width=\textwidth]{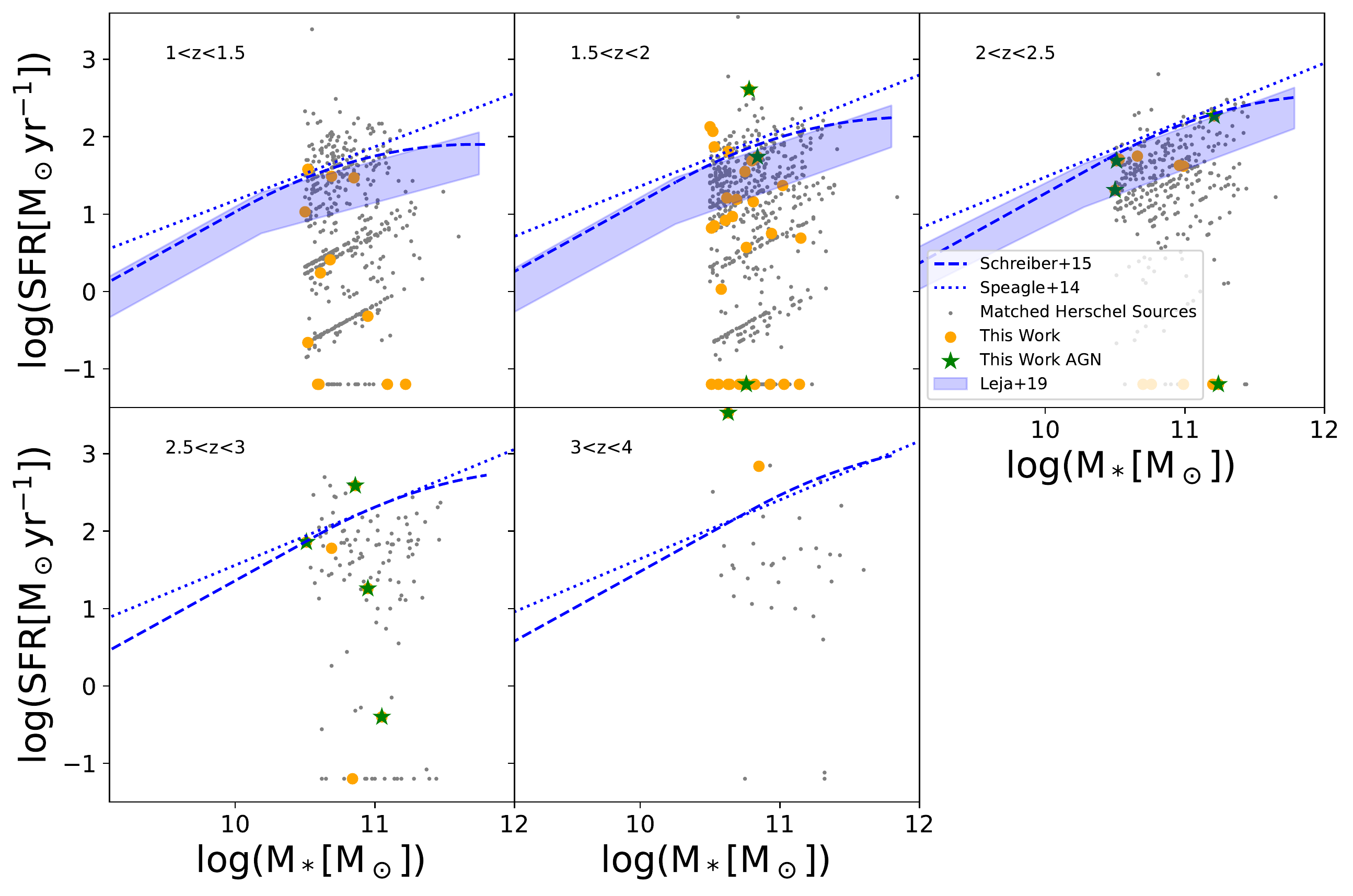}
\caption{Comparison of the SFRs and stellar masses of our sample (orange) with several star-forming main sequence relations from the literature (blue). Green stars indicate AGN. Gray points show the \mbox{\herschel} comparison sample. The SFRs and stellar masses are here derived with FAST++ rather than MAGPHYS as in Figure \ref{fig-main-sequence}. SFRs are clipped below 0.1 \msun yr$^{-1}$ for visual clarity.  \label{figB1}}
\end{figure*}

In Figure \ref{figB1} we show the SFRs and stellar masses of our ultra-dusty sources in comparison to the star-forming main sequence in several redshift bins. Figure 15 shows the results of this comparison when we adopt the SFRs and stellar masses from FAST++ rather than MAGPHYS. As shown in our comparison of the FAST++ and MAGPHYS outputs, the FAST++ SFRs are lower, such that a larger number of sources now fall below the main sequence. We believe the MAGPHYS SFRs to be more robust due to the inclusion of FIR data in the SED modeling, especially for this particular sample of sources for which nearly all the star formation is obscured.

\section{Comparison of $K_S$ and \hst GALFIT Model Results}
\begin{figure*}
\includegraphics[width=\textwidth]{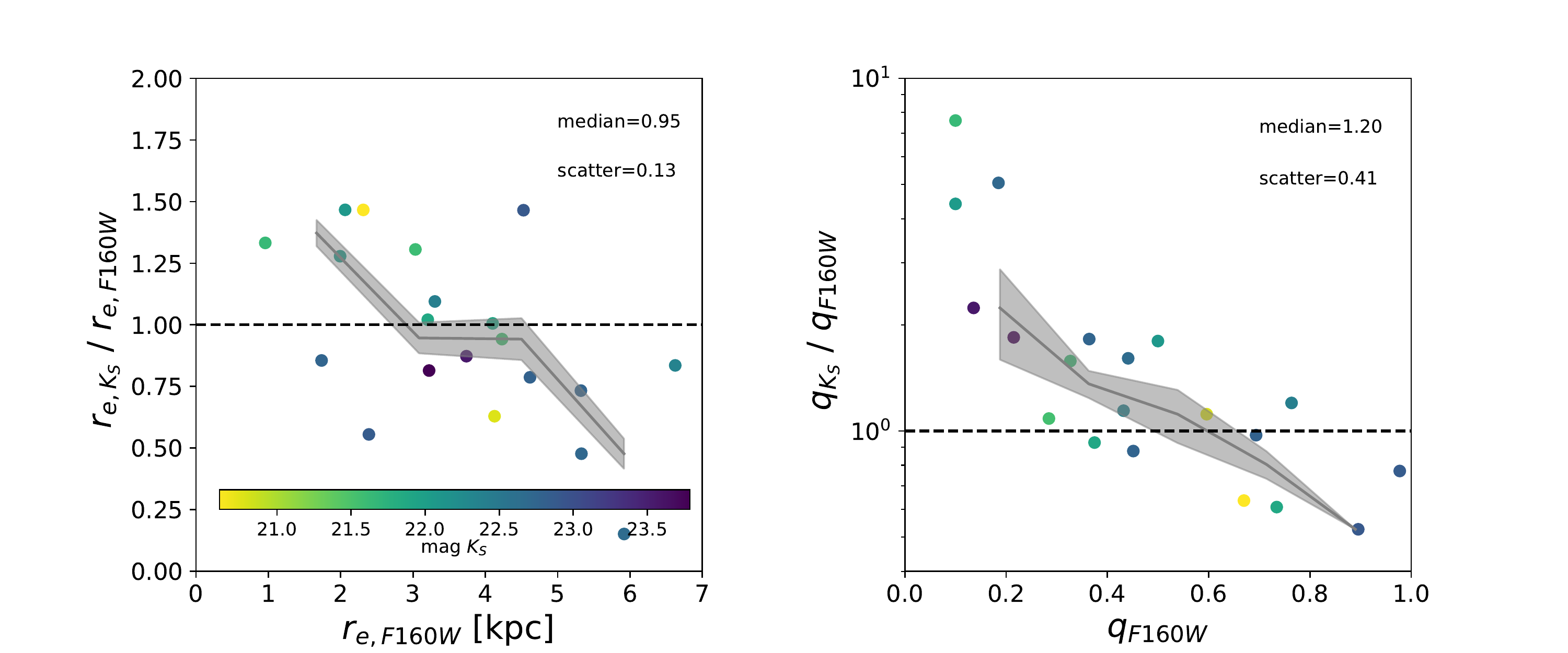}
\caption[width=\textwidth]{Comparison of the $\hst$ $F160W$ and ground-based K$_S$-band sizes (left) and axis ratios (right) measured with GALFIT. The gray curve shows the running mean and scatter. The median offset and scatter for each quantity are listed in each panel.\label{figC1}}
\end{figure*}

Figure \ref{figC1} shows the comparison of the $\hst$ F160W and ground-based K$_S$-band sizes and axis ratios measured with GALFIT when both are available. We find no significant systematic offset in the sizes with a $1\sigma$ scatter of $13\%$. We do observe a systematic trend in the difference between the axis ratios measured in the two bands, such that sources with low axis ratios in F160W have higher axis ratios in the K$_S$-band. We identify two possible reasons for this discrepancy. First, the $\hst$ imaging is sampling a shorter wavelength, meaning both that the imaging is more sensitive to the extreme dust emission and that the sampled stellar population is younger. Second, the sizes measured in the K$_S$-band approach the size of the PSF for our smaller sources, meaning that the short axis may be puffed out. Thus it is not surprising to observe a more elongated shape in the higher resolution imaging. Ultimately, we choose to adopt the K$_S$-band measurements so that we can investigate morphological information for the whole sample.

\section{Alternative Selection of Ultra-Dusty Sources}
\begin{figure}
\includegraphics[width=\columnwidth]{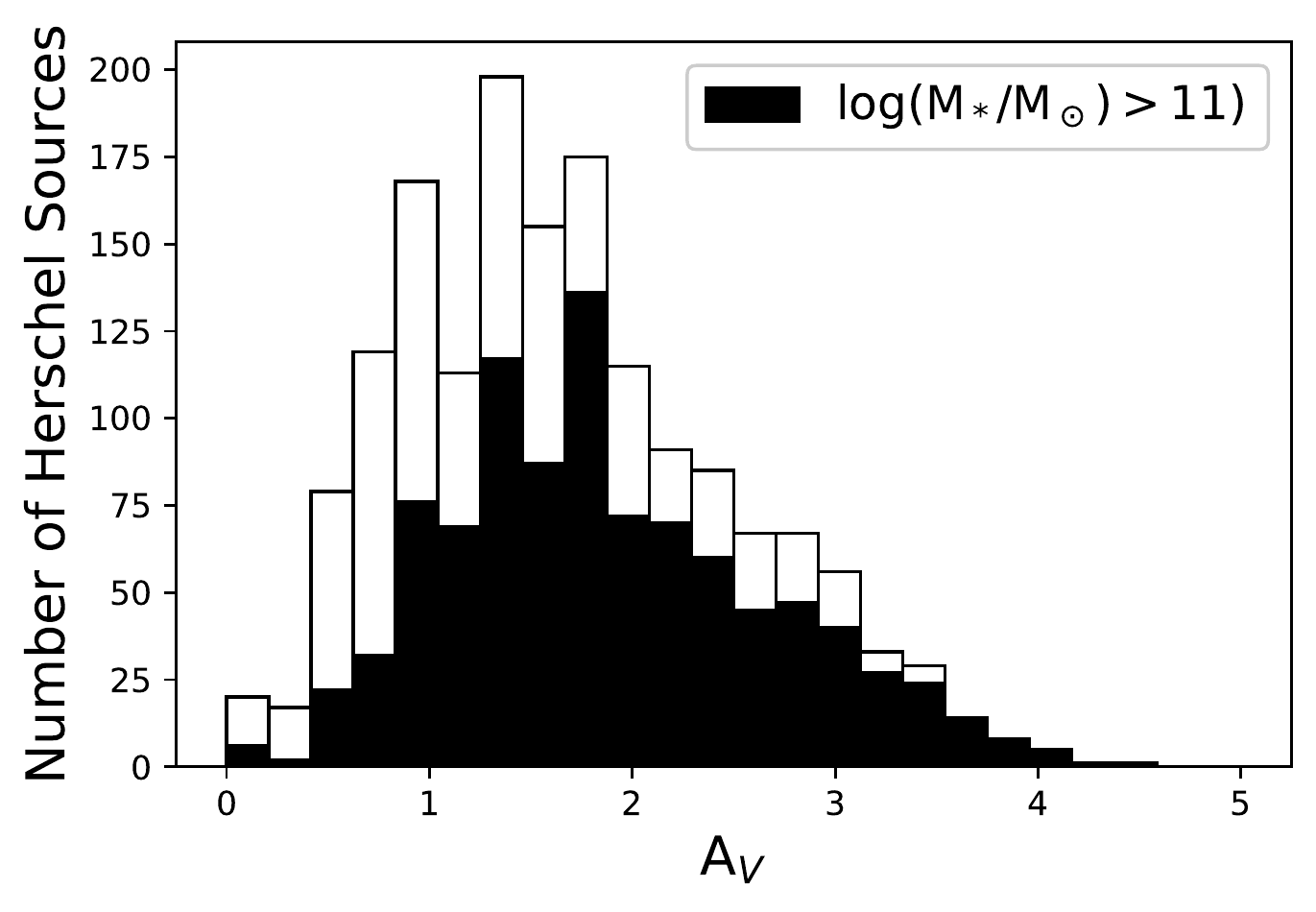}
\caption[width=\columnwidth]{Distribution of A$_V$ for our full \textit{Herschel} comparison sample. Filled histograms indicate sources with log(M$_*/\msun) \ge 11$ from the MAGPHYS modeling. \label{figD1}}
\end{figure}

One of the primary goals of this study is to test whether we can reliably identify heavily-obscured galaxies when FIR data are not available, but for completeness, we examine here a selection using the \textit{Herschel} data. From our \mbox{\herschel} comparison sample, if we instead use the SED fitting results from MAGPHYS to perform the sample selection ($A_V>3$ and log(M$_*/\msun) \ge 11$ to account for the difference in stellar mass estimates from FAST++ and MAGPHYS), we find 107 sources ($\sim 7\%$ of the \textit{Herschel} comparison sample). Figure \mbox{\ref{figD1}} shows the A$_V$ distribution of the full \textit{Herschel} comparison sample. Filled histograms show sources which also meet the mass cut. 25 of these 107 are in common with our sample of 65. We can therefore construct a comparably sized sample of heavily obscured sources by beginning with a FIR selection. This reinforces our conclusion that our sample overlaps with FIR-selected galaxy samples, but is not encompassed by them. Or in other terms, a given level of dust attenuation in rest-frame UV-optical bands corresponds to a range in dust emission levels.

\bsp	
\label{lastpage}
\end{document}